\documentclass[oneside,11pt]{article}

\usepackage[utf8]{inputenc}
\usepackage[T1]{fontenc}

\usepackage{amsthm}
\usepackage{amsmath}
\usepackage{amsfonts}
\usepackage{amssymb}

\usepackage{graphicx}
\usepackage{graphbox}
\usepackage{float}
\usepackage{booktabs}
\usepackage{multirow}
\usepackage{rotating}
\usepackage{array}
\usepackage{xcolor}
\usepackage{subcaption}
\usepackage[ruled]{algorithm2e}

\newcolumntype{C}[1]{>{\centering\arraybackslash}p{#1}}
\newcolumntype{R}[1]{>{\raggedleft\let\newline\\\arraybackslash\hspace{0pt}}m{#1}}
\newcolumntype{L}[1]{>{\raggedright\let\newline\\\arraybackslash\hspace{0pt}}m{#1}}

\usepackage{breakcites}

\usepackage{nowidow}

\usepackage{enumitem}

\usepackage[top=1.5cm,bottom=2cm,right=2.5cm,left=2.5cm]{geometry}
\usepackage{titling}

\usepackage{natbib}

\usepackage{ifplatform}

\usepackage{textcomp}
\usepackage{bbm}

\usepackage{comment}

\ifwindows
\fi

\allowdisplaybreaks

\DeclareFontFamily{OT1}{pzc}{}
\DeclareFontShape{OT1}{pzc}{m}{it}{<-> s * [1.10] pzcmi7t}{}
\DeclareMathAlphabet{\mathpzc}{OT1}{pzc}{m}{it}

\newcommand{\pct}{\%}

\usepackage{multicol}       
\DeclareFontFamily{T1}{cmtt}{\hyphenchar\font=45\relax}

\usepackage[pdftex,final,bookmarksnumbered,bookmarksopen=false,breaklinks,colorlinks]{hyperref}
\hypersetup{
	final,
	bookmarksnumbered=true,
	bookmarksopen=true,
	bookmarksopenlevel=0,
	unicode=false,
	pdftoolbar=true,
	pdfmenubar=true,
	pdffitwindow=false,
	pdftitle={Soprano voices in opera seria: a corpus-based inquiry into eighteenth-century vocal types},
	pdfdisplaydoctitle=true,
	pdftoolbar=true,
	pdfmenubar=true,
	pdflang={English},
	pdfauthor={Ana Llorens, Eduardo Garcia-Portugues, Carlos Vaquero, Alvaro Torrente},
	pdfsubject={arXiv paper},
	pdfcreator={Eduardo Garcia-Portugues},
	pdfproducer={Eduardo Garcia-Portugues},
	pdfkeywords={Castrato}{Corpus study}{Gender}{Logistic regression}{Opera seria}{Soprano}{Statistical learning}{Vocal range},
	pdfnewwindow=true,
	breaklinks=true,
	hidelinks,
	linkcolor=black,
	citecolor=black,
	filecolor=magenta,
	urlcolor=cyan
}

\newif\ifmain
\maintrue
\newif\ifsupplement
\supplementtrue

\newif\iffigstabs
\figstabstrue

\begin{document}

\ifmain

\title{Soprano voices in opera seria: a corpus-based inquiry into eighteenth-century vocal types}
\setlength{\droptitle}{-1cm}
\predate{}%
\postdate{}%
\date{}

\author{Ana Llorens$^{1,4}$, Eduardo Garc\'ia-Portugu\'es$^{2}$, Carlos Vaquero$^{3}$, and \'Alvaro Torrente$^{1,3}$}
\footnotetext[1]{Department of Musicology, Universidad Complutense de Madrid (Spain).}
\footnotetext[2]{Department of Statistics, Universidad Carlos III de Madrid (Spain).}
\footnotetext[3]{Instituto Complutense de Ciencias Musicales, Madrid (Spain).}
\footnotetext[4]{Corresponding author. e-mail: \href{mailto:allorens@ucm.es}{allorens@ucm.es}.}
\maketitle

\begin{abstract}
	Eighteenth-century Italian opera seria was dominated by soprano voices. Although masculine roles were typically performed by castrati and feminine roles by women, cross-casting was common, producing four soprano configurations in which performers of either sex could portray characters of either gender. We hypothesize that composers tailored their writing both to the singers premiering their arias and to the characters' gender. Using statistical models built on approximately 1,700 arias, we analyze relationships between vocal typology, characters' gender, and stylistic choices, comparing supervised learning methods and selecting ridge logistic regression as our primary model, whose performance is competitive with that of the best-performing alternative while allowing for interpretation and uncertainty quantification of its coefficients. Results reveal an asymmetry: character gender is consistently encoded in the writing, while the singer's sex is barely recoverable. Masculine-character arias feature large melodic leaps, intervallic variability, and wider ranges; feminine-character arias favor higher maximum pitches, minor intervals, and melodic stability. Female singers sang higher pitches than male sopranos, but this correlates more with character portrayal than with physiology.
\end{abstract}
\begin{flushleft}
\end{flushleft}

\section{Introduction}
\label{sec:intro}

Italian opera seria---the heroic genre that dominated European operatic stages for most of the eighteenth century---offers exceptionally favorable conditions for corpus-based research. Its production system was intensely conventional: a small canon of dramatic texts---around a dozen of the twenty-six \textit{drammi per musica} of Pietro Metastasio (1698--1782)---was set to music over a thousand times by no fewer than four hundred composers throughout more than a century. For the five most popular dramas analyzed in this study, 554 distinct musical settings have been documented, of which 208---roughly two in five---survive \citep{llorens2024}. Each new production of a given libretto retained the plot, the characters, and the poetic text of most arias, while the music was normally composed anew. This practice of systematic re-composition amounts to a natural experiment rarely available in music history: the dramatic variables are held essentially constant while the musical setting varies, so that compositional choices can be isolated and compared at scale.

From the opening of the first public theatres in Venice, opera was distinctly characterized by a preference for high-pitched voices \citep[p.~176]{Glixon}, a tendency that reached its apex in opera seria. The genre's unwritten rules---the \textit{convenienze teatrali}---prescribed casts of six or seven characters organized in a strict hierarchy: three principals, a second couple, and one or two lesser roles (\citealp[pp.~7--8]{feldman2007opera}; \citealp{TorrenteDominguez2025}). This hierarchy was articulated along two largely independent axes. The amorous-heroic axis comprised the leading couple---the \textit{primo uomo} and the \textit{prima donna}---and their secondary counterparts---the \textit{secondo uomo} and the \textit{seconda donna}---all of them high voices, soprano or alto, notated in C1 and C3 clefs, regardless of the character's gender. The axis of authority was embodied by the third principal, the \textit{tenore}, typically a senior figure---a monarch or a father---usually composed for natural male voice (a tenor or, occasionally, a bass) \citep{TorrenteDominguez2025}. Register thus encoded dramatic function and decorum---youthful heroism and amorous passion versus age and institutional authority---rather than gender alone \citep{freitas}. Soprano roles predominated among the \textit{primi} pair (95\pct), compared with \textit{secondi} (84\pct; see \citealp{didonedatabase}).

Within this heroic register, two kinds of singers were available: women and castrati---male sopranos castrated before puberty to preserve their unbroken treble, a practice rooted in sixteenth-century church music that reached its artistic zenith on the eighteenth-century operatic stage: ``The entire classical foundation of virtuosic solo singing in the West, [...] owes its existence to the musical traditions and practices of castrati'' \citep[pp.~xi--xii]{feldman2015castrato}. Evidence of castrated singers (``capones'') in cathedral chapels is documented in Spain as early as 1506, half a century before Pope Paul IV (1555--1559) forbade women from singing in churches---an action that contributed to the rise of the castrati. See \citet[pp.~47--52]{medina2001atributos} and \citet{Rosselli_1988}. Over the eighteenth century, the demands on the castrati's virtuosity grew steadily, expanding both their tessitura (the region of the range in which a part predominantly lies) and the variety and complexity of their coloratura (rapid, florid passage-work). Although castrati and female sopranos commanded essentially the same ranges, contemporary accounts insist on differences in their sound: an exceptional vocal power, commonly linked in the literature to the castrati's purportedly expanded thoracic capacity, which allegedly allowed greater control of sound intensity and superior breath support for extended phrases (\citealp[pp.~79--132]{feldman2015castrato}; \citealp[pp.~21--24]{seedorf2015}), and a particularly brilliant and penetrating timbre---``an irreducible acoustic difference between castrati and all other singers: a singer's formant in the prevalent pitch area of f'/g' through c'' to e'' achieved without `pushing' the voice'' (\citealp[p.~99]{feldman2015castrato}; \citealp{Sundberg_sopranos}). The proverbial ``long breath'' of the castrati is, moreover, documented in the parts written for them, which frequently demand long-held notes and extended coloratura passages \citep[p.~24]{seedorf2015}. These physiological and acoustic differences---which no recording preserves---motivate one of the central hypotheses of this study: that composers may have tailored their vocal writing to the specific type of soprano expected to sing it.

In casting, the customary alignment matched masculine characters with castrati and feminine characters with women \citep{freitas}. However, cross-casting was not unusual, particularly in the Papal States---as well as in Portugal---where women were banned from the stage, necessitating castrati in the feminine roles. Elsewhere, women not infrequently performed masculine roles, following the practical requirements of each production and the abilities of the available singers; women performing masculine roles were more common in lesser theatres which could not afford the high fees of castrati \citep[pp.~41--42]{seedorf2015}. Thus, within the soprano settings, four different options are found in the opera seria tradition: coincidence of character gender and biological sex of the singers---male singer performing a masculine character, or female singer performing a feminine character---was most common, though the other two options---male singer performing a feminine character, or female singer performing a masculine character---were also relatively frequent. (For the sake of clarity, we use \emph{masculine} and \emph{feminine} to refer to the characters' gender, and \emph{male}, and \emph{female} to indicate the singers' biological sex.) This interchangeability has a consequence that defines the scope of our study: within the soprano register, the type of voice by itself carried virtually no information about a character's gender, since performers of either sex could---and did---portray characters of either gender. If composers marked gender musically, the marking must therefore reside in the vocal writing itself---in the intervals, contours, and textures of the arias---rather than in the type of singer delivering them. Cross-casting must be distinguished from diegetic cross-dressing, in which a character disguises themselves as another gender within the plot. The latter was rare in opera seria: only three explicit cases appear in Metastasio's dramatic corpus: Emira disguised as Idaspe in \textit{Siroe re di Persia} (1726), Semiramide as Nino in \textit{Semiramide riconosciuta} (1729), and Achille as Pirra in \textit{Achille in Sciro} (1736). See \citet{Heller}.

Eighteenth-century operatic decorum suggests what such marking might look like: heroic masculinity is intuitively linked with wide leaps and martial, trumpet-like figuration, femininity with higher tessitura and smoother, more conjunct lines. These expectations differ in their grounding, however. Those concerning voice-specific writing have documented support in the long-held notes and extended coloratura passages already mentioned \citep{seedorf2015}. Those concerning gendered style, by contrast, circulate as tacit assumptions: seldom stated explicitly in the scholarly literature, never quantified. At the opposite pole, it has even been argued explicitly that vocal register carried no dramatic connotation whatsoever---that the conventions of the genre were ``a dramatic custom devoid of specific connotations'' (\citealp[p.~102]{moindrot1993}; similarly \citealp{keyser1987}; both discussed in \citealp{freitas}). Scholarship has also long documented the flexibility of the casting system and the functional interchangeability of castrati and female sopranos in leading roles \citep{charton2012,feldman2015castrato, Heller, seedorf2015}. Treating these competing positions as explicit, falsifiable hypotheses allows us to ask not only whether gendered conventions are detectable in the notated music, but also where compositional practice departs from them---and it is in these departures, as we shall see, that the corpus proves most informative.

One further convention bears on the interpretation of our data. The dominant aria form of the period was the da capo aria, a tripartite A--B--A' structure in which the return of the first section was expected to be freely embellished by the singer with improvised ornaments and cadenzas---the moment in which singers, the true stars of the eighteenth-century operatic system, completed the work in performance \citep{Poriss}. Because such ornamentation was not notated, the scores analyzed here document the composer's written text rather than the sounding performance: our features measure compositional choices, not the performative layer added upon them. This distinction will prove relevant when interpreting the results.

This study investigates how vocal writing in Italian opera seria reflects conventional vocal classifications based on both the biological sex of performers (male/female) and the gender of dramatic characters (masculine/feminine). We hypothesize that composers adapted their writing to both singers' characteristics and characters' genders. Furthermore, we propose that the unique physiological qualities of male (i.e., castrati) versus female sopranos influenced the specific characteristics of their vocal lines. To examine these questions, we adopt a data-driven approach, applying statistical learning methods to analyze patterns in a large corpus of arias. Our analysis comprises a set of case studies in which we apply classification models to multiple subsets of musical features, specifically designed to capture vocal dimensions relevant to soprano singing in a mid- and low-level~\mbox{representation.}

We develop our investigation around the following questions:
\begin{itemize}
    \item To what extent can the gender of operatic characters, as delineated by the libretto, be inferred from the musical characteristics of their arias?
    \item When arias are performed by male versus female sopranos, does the classification of the characters' gender differ? If so, in what specific ways does this distinction manifest?
    \item Are there systematic musical distinctions between arias sung by male sopranos and those sung by female sopranos?
\end{itemize}

To respond to these questions, we organize the investigation into three separate case studies:
\begin{itemize}
    \item \textbf{Case study 1}: We investigate whether composers differentiated their vocal writing according to the dramatic gender of characters, irrespective of the performer's biological sex, using the full corpus of soprano arias.
    \item \textbf{Case study 2}: We explore whether the musical features used to portray characters' dramatic gender are influenced by the biological sex of the singers, focusing on arias where the character's dramatic gender aligns with the singer's biological sex.
    \item \textbf{Case study 3}: Since composers frequently wrote with specific singers in mind, we investigate whether they tailored their vocal lines to the soprano singers' biological sex, that is, whether the performers were female sopranos or male castrati.
\end{itemize}
Through this approach, we explore how the intersections between vocal type and gender are encoded in the musical fabric of opera seria, revealing a more nuanced and dynamic compositional landscape than its formal conventions might initially suggest.

\section{Materials and methods}
\label{sec:methods}

\subsection{Corpus and data sources}
\label{Dataset}

The dataset used in this study consists of symbolic and metadata features extracted from 1,682 soprano arias---by far the most abundant in this repertoire, and mutually comparable in range---from operas based on Pietro Metastasio's five most popular \textit{drammi per musica}. These are \textit{Didone abbandonata} (1724), \textit{Alessandro nell'Indie} (1730), \textit{Artaserse} (1730), \textit{Adriano in Siria} (1732), and \textit{Demofoonte} (1733). The corpus is broad in both authorship and time: the arias are due to 99 identified composers (18 further arias remain anonymous), the dated settings span the years 1724--1810, and 303 different singers are documented as having premiered them. Coverage across the five dramas is uneven, reflecting the surviving soprano settings available to us at this moment: \textit{Alessandro nell'Indie} contributes 486 arias, \textit{Didone abbandonata} 438, \textit{Demofoonte} 411, and \textit{Artaserse} 336, whereas \textit{Adriano in Siria} is only marginally represented, with 8 arias for two of its characters (Emirena and Sabina); three further arias are settings of texts imported from other Metastasio dramas into productions of these operas, one of them for the character of Tamiri (see Table~\ref{tab:characters_singers_ratios}).

Depending on the case study, we work with a specific portion of the dataset (see Table~\ref{tab:distribution_arias}), namely:
\begin{itemize}
    \item \textbf{Case study 1}: The dataset for this case study is the full corpus, containing 805 arias for feminine soprano characters and 877 arias for masculine soprano characters. We therefore work with two reasonably balanced classes, as is the case with the following case studies as well. The target label for this case study is the characters' gender for each aria.
    \item \textbf{Case study 2}: The dataset used in this case study is restricted to arias for feminine characters performed by only female sopranos, as well as arias for masculine characters performed by only male sopranos. The target label is, as in Case study 1, the gender of the character sung, with 670 arias for masculine characters sung by male sopranos, and 569 arias for feminine characters sung by female sopranos. As shown in Table~\ref{tab:distribution_arias}, the dataset available for this case study is smaller than that of Case study 1, as not all libretti include records of the singers who premiered the arias.
    \item \textbf{Case study 3}: The dataset for this case study contains arias for both masculine and feminine characters and singers, for which we know the name of the singer who premiered them. Specifically, the number of arias premiered by male sopranos is 793, and by female sopranos, 643. The target label for this case study is the inferred biological sex of the singers premiering each aria.
\end{itemize}

All scores were provided by the Didone project under an NDA agreement (see \citealp{didoneproject}). They are available for consultation at \citet{didonedatabase}. Having all arias in MusicXML format, we extracted symbolic features using the Python library \texttt{musif} \citep{llorens2023musif}. The complete list of features used in this study is given in Section~\ref{app:features_list}. Given our focus on composition for the vocal parts, from the initial \texttt{musif} extraction, we selected those concerning the soprano voice and the properties associated with them. The extraction done with \texttt{musif} involved global descriptors of the score, without capturing sequential aspects within. The deposited dataset \citep{Zenododataset} retains the 567 \texttt{musif} features that describe the principal soprano part. From these, we discarded by name the windowed and sequential descriptors, the raw counts and sums, the per-interval-type columns, the scale-degree distributions, and the note-name and redundant categorical interval descriptors, retaining the 86 features listed in Section~\ref{app:features_list}, which were then preprocessed as described in Section~\ref{sec:preproc}.

\begin{table}[!ht]
\centering
\caption{\small Ratio of female vs. male singers for each character in the dataset.
Female ratio and Male ratio give the proportion of arias premiered by female and by male sopranos,
computed over the arias with a known singer. Known singer gives the number of such arias, and
Total arias the number of arias available for the character within the corpus used in this study.
Dashes mark the single character with no arias with a known singer, for which the ratios are undefined.}
\label{tab:characters_singers_ratios}
\small
\begin{tabular}{llcccc}
\toprule
\textbf{Character} & \textbf{Gender} & \textbf{Female ratio} & \textbf{Male ratio} & \textbf{Known singer} & \textbf{Total arias} \\ \midrule
Poro & Masculine & 0.04 & 0.96 & 133 & 157 \\
Didone & Feminine & 0.82 & 0.18 & 132 & 152 \\
Cleofide & Feminine & 0.80 & 0.20 & 106 & 121 \\
Enea & Masculine & 0.11 & 0.89 & 103 & 116 \\
Timante & Masculine & 0.12 & 0.88 & 96 & 117 \\
Dircea & Feminine & 0.87 & 0.13 & 95 & 125 \\
Arbace & Masculine & 0.01 & 0.99 & 94 & 101 \\
Mandane & Feminine & 0.82 & 0.18 & 85 & 89 \\
Erissena & Feminine & 0.80 & 0.20 & 79 & 94 \\
Selene & Feminine & 0.82 & 0.18 & 76 & 89 \\
Creusa & Feminine & 0.86 & 0.14 & 59 & 73 \\
Artaserse & Masculine & 0.04 & 0.96 & 55 & 60 \\
Gandarte & Masculine & 0.06 & 0.94 & 53 & 63 \\
Semira & Feminine & 0.75 & 0.25 & 52 & 53 \\
Araspe & Masculine & 0.15 & 0.85 & 47 & 52 \\
Cherinto & Masculine & 0.15 & 0.85 & 46 & 63 \\
Megabise & Masculine & 0.36 & 0.64 & 25 & 25 \\
Alessandro & Masculine & 0.00 & 1.00 & 23 & 28 \\
Timagene & Masculine & 0.21 & 0.79 & 19 & 23 \\
Adrasto & Masculine & 0.54 & 0.46 & 13 & 18 \\
Iarba & Masculine & 0.00 & 1.00 & 11 & 14 \\
Osmida & Masculine & 0.25 & 0.75 & 8 & 15 \\
Artabano & Masculine & 0.00 & 1.00 & 8 & 10 \\
Demofoonte & Masculine & 0.50 & 0.50 & 6 & 10 \\
Emirena & Feminine & 1.00 & 0.00 & 5 & 5 \\
Matusio & Masculine & 0.25 & 0.75 & 4 & 5 \\
Sabina & Feminine & 1.00 & 0.00 & 3 & 3 \\
Tamiri & Feminine & -- & -- & 0 & 1 \\
\bottomrule
\end{tabular}
\end{table}

\begin{table}[!ht]
\centering
\caption{\small Distribution of arias by characters' gender and singers' sex.}
\label{tab:distribution_arias}
\small
\begin{tabular}{lcccc}
\toprule
 & \textbf{Total} & \textbf{Male singers} & \textbf{Female singers} & \textbf{Unknown singers} \\ \midrule
Masculine roles & 877 & 670 & 74 & 133 \\
Feminine roles & 805 & 123 & 569 & 113 \\
\bottomrule
\end{tabular}
\end{table}

Metadata features were extracted from the available libretti. The gender of the characters in each drama is clearly specified. Furthermore, the libretti of the premiere of each musical setting often include the names of the singers performing each role. Metadata were obtained from \citet{corago} and \citet{didonedatabase}. The singers' biological sex was inferred by identifying whether their given names correspond to traditionally male or female Italian names. This method is susceptible to inaccuracies, as it does not account for any additional documentation concerning a singer's gender identity beyond the name registered in the libretti.

Table~\ref{tab:characters_singers_ratios} shows the number of arias available within our corpus for each character, together with the number of arias with libretto-based information about the singers premiering them, and the ratio distribution of female and male soprano singers by aria. The ratios independently reproduce, at scale, the documented stratification of casting practice: the castrato remained the default option for the \textit{primo uomo}, whereas lesser roles were shared freely between castrati and women \citep[pp.~38--41]{seedorf2015}. The aria for Tamiri, originally a character in Metastasio's \textit{Semiramide}, was inserted into Mattia Vento's version of \textit{Demofoonte} (1765). Moreover, some of the main masculine characters, such as Alessandro in \textit{Alessandro nell'Indie}, were originally crafted for a tenor---and sometimes a bass---voice. See \citet{TorrenteDominguez2025}.

As a preliminary exploration of the corpus, Figure~\ref{fig:male_female_ranges} illustrates the lower and upper vocal ranges of the dataset arias that have been registered as premiered by female or male sopranos. Arias composed for male sopranos tended towards slightly lower registers overall, yet most of the notes overlap around the same pitches, which implies that the classification of the singer's sex from these features alone is non-trivial. In the most extreme examples of lowest register usage in our study, we find J.~A.~Hasse's ``Prudente mi chiedi?'' and ``Se ardire e speranza'' (from \textit{Demofoonte}, 1748), in which the character of Timante sings, respectively, an E$\flat$3 and E3. And for the highest notes, we find M.~Mortellari's ``Tuona il cielo incalza il vento'' (from \textit{Alessandro nell'Indie}, 1778) and G.~Colla's ``Va lusingando amore'' (from \textit{Didone abbandonata}, 1773), in which Cleofide and Didone respectively sing a G6.

\begin{figure}[!h]
\centering
\includegraphics[width=1.0\linewidth]{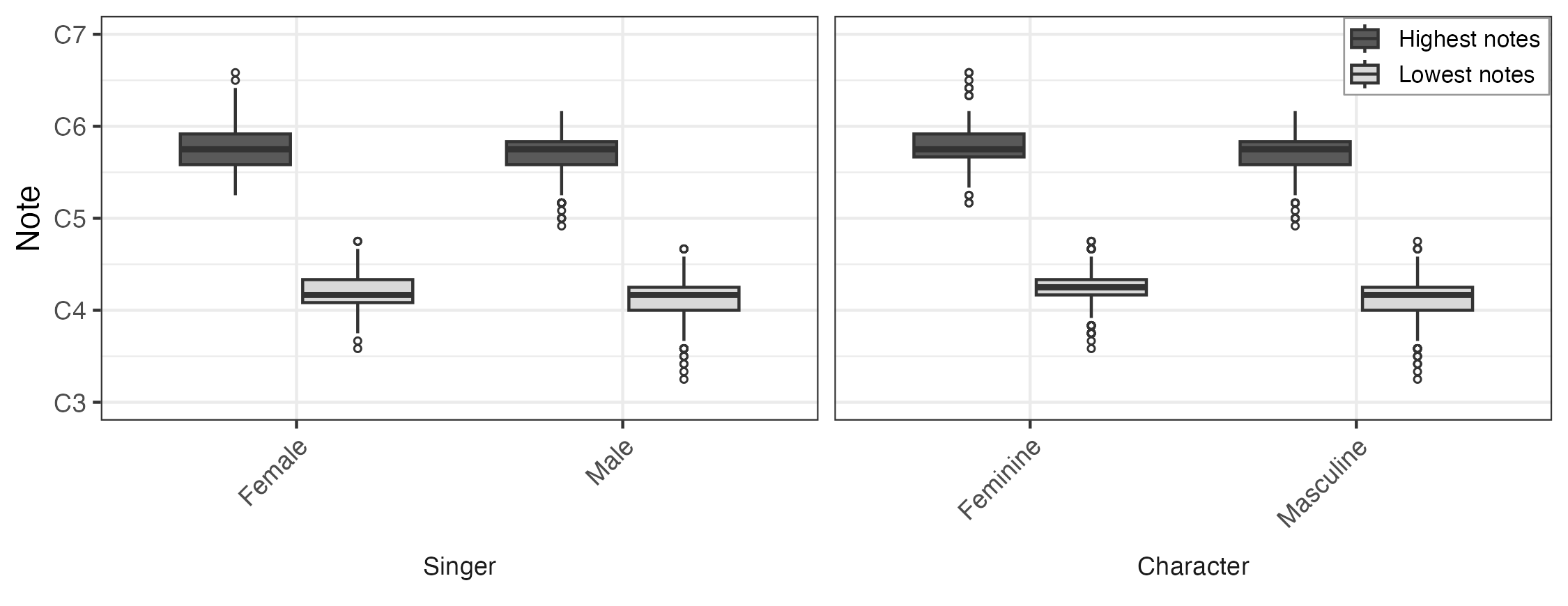}
\caption{\small Vocal ranges by singer sex and by character gender.
Boxplots of the highest and lowest notes of the arias: by the premiering singer's sex for the 1,436 arias with a known singer (left), and by character gender for all 1,682 arias (right).}
\label{fig:male_female_ranges}
\end{figure}

\subsection{Experimental setup}
\label{ml_methods}

The experimental setup was framed as a set of binary classification problems, where the target variable corresponded to either the gender of the character or the biological sex of the singer. Before model training, the dataset was randomly partitioned into training and testing sets, stratified by the response so that both sets preserve the class balance of the full dataset, comprising 70\pct{} and 30\pct{} of the data, respectively, for each case study; the comparatively large test fraction stabilizes the held-out model comparisons. In Case study 1, 1,176 arias were used to train the model through cross-validation, while 506 were reserved for testing. For Case studies 2 and 3, the splits were 866 and 373 arias, and 1,005 and 431 arias, respectively.

To address the classification tasks, we considered three well-established methods---random forest (as classifier) \citep{breiman2001random}, logistic regression with lasso penalization, and logistic regression with ridge penalization \citep{Hastie2015}---and compared their performance before selecting the most suitable method for the case studies presented here. Across all case studies, $Y=0$ encodes ``female/feminine'' and $Y=1$ encodes ``male/masculine''.

\subsection{Data preprocessing}
\label{sec:preproc}

Prior to fitting and evaluating the alternative statistical methods, a preprocessing recipe was applied to remove noise and mitigate multicollinearity among features, thereby enhancing the reliability and performance of the classification algorithms. In particular, predictors with zero or near-zero variance, high proportions of missing values ($>5\pct$), or high pairwise correlations ($>0.8$) were removed. Remaining missing values were imputed using $k$-nearest neighbors ($k = 5$). Rare factor levels ($<10\pct$ frequency) were collapsed into an ``other'' category, and linear dependencies among the numeric predictors were removed to avoid multicollinearity. Categorical predictors were subsequently encoded using one-hot encoding, with the ``other'' category removed to serve as the reference level. Following dummy creation, zero- and near-zero variance predictors as well as linear dependencies were re-evaluated and removed and, as a final step, all predictors were centered and scaled, so that every coefficient of the fitted models refers to a standardized predictor. For the model comparison reported in Section~\ref{sec:modcomp}, all preprocessing steps were trained on the training dataset; for the final models fitted on the whole dataset of each case study, they were trained on all available arias. The preprocessing was done with the \texttt{recipes} package \citep{recipes}.

The preprocessing step allowed us to trim down the number of features from 86 to 34 in Case study 1, to 35 in Case study 2, and to 34 in Case study 3 for the final models fitted on all arias; the models trained on the training subsets retain the same numbers of predictors, although the data-driven steps resolve slightly differently and one retained feature differs in each case study.

\subsection{Logistic regression with regularization}
\label{LogReg}

Logistic regression models the conditional probability of the class $Y=1$ as a logistic-function transformation of a linear combination of the covariates $X_1,\ldots,X_p$:
\begin{equation*}
    \mathbb{P}[Y=1|X_1=x_1,\ldots,X_p=x_p]=\mathrm{logistic}(\eta(\boldsymbol{x})),
\end{equation*}
where $\mathrm{logistic}(x):=1/(1+e^{-x})$ and $\eta(\boldsymbol{x})=\beta_0+\beta_1x_1+\cdots+\beta_px_p=\beta_0+\boldsymbol{x}^\top\boldsymbol{\beta}_{-1}$ is the linear predictor. The coefficients $\boldsymbol{\beta}=(\beta_0,\beta_1,\ldots,\beta_p)^\top$ admit a clear interpretation in terms of the conditional odds of $Y=1$ vs. $Y=0$, $\mathbb{P}[Y=1|X_1=x_1,\ldots,X_p=x_p]/\mathbb{P}[Y=0|X_1=x_1,\ldots,X_p=x_p]$. Indeed, for $j=1,\ldots,p$, $e^{\beta_j}$ represents the multiplicative increment of the conditional odds attributed to an increment of one unit in the predictor $X_j$, provided the remaining predictors are held constant. For example, if $\beta_1=0.25$, then $e^{\beta_1}\approx 1.2840$, meaning that the conditional odds of $Y=1$ vs. $Y=0$ increase by $28.40\pct$ when increasing $X_1$ by one unit, thus allowing for a clear quantification of the effects of a predictor on the classification of $Y$.

Regularization in the form of shrinkage is needed to fit high-dimensional logistic regressions. This regularization can be enforced through an elastic net penalty \citep{Hastie2015} that, after the predictors have been standardized, estimates $\boldsymbol{\beta}$ from the sample $\{(\boldsymbol{x}_i,y_i)\}_{i=1}^n$ as
\begin{equation}
    \hat{\boldsymbol{\beta}}_{\lambda}:=\arg\min_{\boldsymbol{\beta}\in\mathbb{R}^{p+1}}\left\{-\ell(\boldsymbol{\beta})+\lambda (\gamma\|\boldsymbol{\beta}_{-1}\|_1+(1-\gamma)\|\boldsymbol{\beta}_{-1}\|_2^2)\right\},\label{eq:eln}
\end{equation}
where
\begin{equation*}
    \ell(\boldsymbol{\beta})=\frac{1}{n}\sum_{i=1}^n\left[y_i\log(\mathrm{logistic}(\eta(\boldsymbol{x}_i)))+(1-y_i)\log(1-\mathrm{logistic}(\eta(\boldsymbol{x}_i)))\right]
\end{equation*}
is the average log-likelihood. Lasso logistic regression is obtained with $\gamma=1$ and has the advantage of zeroing coefficients, hence facilitating the interpretation of the model. Ridge logistic regression uses $\gamma=0$ and imposes a global shrinkage on the coefficients.

Ridge regression has the advantage of allowing for simple asymptotic inference based on $\hat{\boldsymbol{\beta}}_{\lambda}$, unlike the more convoluted selective \citep{Lee2016} and debiasing \citep{VanDeGeer2014} inference approaches for lasso. Indeed, $\hat{\boldsymbol{\beta}}_{\lambda}$ is asymptotically normally distributed with asymptotic covariance \citep[p.~194]{Cessie1992}
\begin{equation}
    \boldsymbol{\Sigma}_{\lambda}:=(\mathbb{X}^\top\boldsymbol{V}\mathbb{X}+
    \boldsymbol{I}_{\lambda})^{-1}(\mathbb{X}^\top\boldsymbol{V}\mathbb{X})(\mathbb{X}^\top\boldsymbol{V}\mathbb{X}+\boldsymbol{I}_{\lambda})^{-1}, \label{eq:asympvar}
\end{equation}
where $\mathbb{X}$ is the $n\times (p+1)$ matrix formed by a column of ones and the entries $x_{ij}$, $\boldsymbol{V}=\mathrm{diag}(V_1,\ldots,V_n)$ with $V_i=\mathrm{logistic}(\eta(\boldsymbol{x}_i))(1-\mathrm{logistic}(\eta(\boldsymbol{x}_i)))$, $i=1,\ldots,n$, and $\boldsymbol{I}_{\lambda}=\mathrm{diag}(0,2n\lambda,\ldots,2n\lambda)$ is a $(p+1)\times (p+1)$ diagonal matrix. The asymptotic expectation of $\hat{\boldsymbol{\beta}}_{\lambda}$ is the population logistic ridge vector of coefficients
\begin{equation}
\begin{split}
    \boldsymbol{\beta}_{\lambda}:=\arg\min_{\boldsymbol{\beta}\in\mathbb{R}^{p+1}}\big\{&-\mathbb{E}_{(\boldsymbol{X},Y)}\left[Y\log(\mathrm{logistic}(\eta(\boldsymbol{X})))+(1-Y)\log(1-\mathrm{logistic}(\eta(\boldsymbol{X})))\right]\\
    &+\lambda \|\boldsymbol{\beta}_{-1}\|_2^2\big\}.
\end{split}
\label{eq:asympbias}
\end{equation}
Let $\hat{\boldsymbol{\Sigma}}_{\lambda}$ denote Eq.~(\ref{eq:asympvar}) with $\boldsymbol{V}$ evaluated at $\hat{\boldsymbol{\beta}}_{\lambda}$. From the previous results, we get the asymptotically valid (fixed-$\lambda$) $100(1-\alpha)\pct$ confidence intervals for the entries of Eq.~(\ref{eq:asympbias}):

\begin{equation}
    \mathrm{CI}_{1-\alpha}(\beta_{\lambda,j})=\left(\hat{\beta}_{\lambda,j}-z_{1-\alpha/2}\sqrt{\hat{\Sigma}_{\lambda,jj}},\hat{\beta}_{\lambda,j}+z_{1-\alpha/2}\sqrt{\hat{\Sigma}_{\lambda,jj}}\right),\quad j=0,\ldots,p,\label{eq:cis}
\end{equation}
with $z_{1-\alpha/2}$ being the $(1-\alpha/2)$-quantile of a standard normal. Since the $p$ slope coefficients are tested simultaneously, we guard against multiplicity through a Bonferroni correction: the reported confidence intervals for $\beta_{\lambda,j}$, $j=1,\ldots,p$, are computed at the $1-\alpha/p$ level, so that they hold jointly with confidence at least $100(1-\alpha)\pct$. We take $\alpha=0.05$ throughout.

To fit Eq.~(\ref{eq:eln}) we used the R package \texttt{glmnet} \citep{glmnet}, using stratified $10$-fold cross-validation to select the tuning parameter $\lambda$ by the \emph{one standard error rule} $\hat{\lambda}_{1\mathrm{SE}}$ that minimizes binomial deviance. This rule chooses the penalty $\lambda$ that yields the simplest model within one standard error margin of the minimizer of the cross-validation loss. The confidence intervals in Eq.~(\ref{eq:cis}) do not account for this data-driven selection of $\lambda$, i.e., they are conditional on $\hat{\lambda}_{1\mathrm{SE}}$. For ridge regression, the correspondence between the parametrization of $\lambda$ in Eq.~(\ref{eq:eln}) and that used in \texttt{glmnet} is $\lambda = \lambda_{\texttt{glmnet}}/2$.

\subsection{Random forest}
\label{RF}

A random forest \citep{breiman2001random} is an ensemble classifier constructed from a collection of $T$ decision trees $\{ h_1, h_2, \dots, h_T \}$ grown independently. Each tree recursively partitions the covariate space $(X_1,\ldots,X_p)$ through binary splits of the form $\{X_j\leq c_j\}$ vs. $\{X_j>c_j\}$, chosen greedily to maximize the decrease in Gini impurity, until a minimum node size is reached. Diversity among the trees is induced by training each tree on a bootstrap resample of $\{(\boldsymbol{x}_i,y_i)\}_{i=1}^n$ and by restricting each split search to a random subset of \texttt{mtry} of the $p$ covariates. For an input $\boldsymbol{x}$, the forest outputs the class-probability estimate $\hat{p}(\boldsymbol{x}) := (1/T)\sum_{t=1}^T 1_{\{ h_t(\boldsymbol{x}) = 1 \}}$, the proportion of trees voting for class $1$, and the class prediction $\hat{y}(\boldsymbol{x}) = 1_{\{\hat{p}(\boldsymbol{x})\geq 0.5\}}$, i.e., the majority vote over the $T$ trees.

We used the \texttt{randomForest} R package \citep{randomForest} to fit the model with $T = 5{,}000$ trees and a minimum node size of one, and the \texttt{caret} package \citep{caret} to select \texttt{mtry} over a grid of five candidate values via cross-validation, with the area under the ROC curve (AUC) computed from $\hat{p}$ as the performance metric. As in logistic regression with shrinkage, we applied stratified $10$-fold cross-validation.

\subsection{Model evaluation}
\label{sec:modcomp}

To compare the performance of the models obtained with logistic regression and random forest, we use the Accuracy (Acc), that is, the proportion of correct classifications of a model. Mathematically, if $Y_1,\ldots, Y_{n_{\text{test}}}$ are the labels of the testing dataset and $\hat{Y}_1,\ldots, \hat{Y}_{n_{\text{test}}}$ are their predictions with model $M$, then $\mathrm{Acc}(M):=(1/n_{\text{test}}) \sum\limits_{i=1}^{n_{\text{test}}} 1_{\{Y_i=\hat{Y}_i\}}$. We consider this simple and interpretable metric given the highly balanced nature of the three case studies, where the proportions of the majority classes are $877/1,682\approx 0.52$, $670/1,239\approx 0.54$, and $793/1,436\approx0.55$, respectively. For completeness, we also report the AUC and the $\mathcal{F}_1$ macro score \citep{Opitz2019}. The latter is the symmetric version of the $F_1$ score that averages the two $F_1$ scores where each class acts as the positive class.

Acknowledging the variability in model comparison, we used bootstrap resampling to assess the significance of the difference in accuracies between the trained models. Precisely, let $M_j$, $j=1,2$, be two fitted models, each yielding classifications $\hat{Y}^{(M_j)}_1, \ldots, \hat{Y}^{(M_j)}_{n_\text{test}}$. These two samples generate the accuracies $\mathrm{Acc}(M_1)$ and $\mathrm{Acc}(M_2)$, and the difference $\Delta (M_1,M_2):=\mathrm{Acc}(M_1)-\mathrm{Acc}(M_2)$. A percentile bootstrap $100(1-\alpha)\pct$-confidence interval for the (population) difference in accuracy is obtained by:
\begin{enumerate}
    \item Drawing $b=1,\ldots,B$ bootstrap samples: for each $b$, indices $i_1^{*b},\ldots,i_{n_\text{test}}^{*b}$ are drawn with replacement from $\{1,\ldots,n_\text{test}\}$, yielding the resampled triples $\big\{\big(Y_{i_k^{*b}},\allowbreak\hat{Y}^{(M_1)}_{i_k^{*b}},\allowbreak\hat{Y}^{(M_2)}_{i_k^{*b}}\big)\big\}_{k=1}^{n_\text{test}}$, so that each label is resampled jointly with its two predictions.
    \item Setting $\{\Delta^{*b} (M_1,M_2):=\mathrm{Acc}^{*b}(M_1)-\mathrm{Acc}^{*b}(M_2)\}_{b=1}^{B}$.
    \item Computing the percentile confidence interval
    \begin{equation*}
    	\mathrm{CI} (M_1,M_2)=\Big(\Delta^{*(\lceil (B+1)\alpha/2 \rceil)}(M_1,M_2),\Delta^{*(\lfloor (B+1)(1-\alpha/2)\rfloor)}(M_1,M_2)\Big),
    \end{equation*}
    where $\Delta^{*(b)}(M_1,M_2)$ stands for the $b$th ordered bootstrap accuracy difference.
\end{enumerate}
Note the above (simple) bootstrap procedure does not retrain the models $M_1$ and $M_2$.

If zero belongs to the confidence interval $\mathrm{CI}(M_1,M_2)$, then the difference in accuracy between $M_1$ and $M_2$ is not significant at significance level $\alpha$: we do not find evidence of a performance difference between the two models after factoring out the uncertainty in the testing set variability, given the fitted models. More precisely, we can investigate whether model $M_1$ is significantly \emph{better} than model $M_2$ with a one-sided hypothesis test on the (population) accuracies: $H_0\colon \mathrm{Acc}(M_1)=\mathrm{Acc}(M_2)$ vs. $H_1\colon \mathrm{Acc}(M_1)>\mathrm{Acc}(M_2)$. A bootstrap $p$-value, denoted $p_{M_1>M_2}$, is obtained by inverting the one-sided percentile bootstrap confidence interval, as a side-product of the above algorithm:
\begin{equation*}
    p_{M_1>M_2}:=\frac{1}{B+1}\left\{1+\sum_{b=1}^B1_{\{\Delta^{*b}(M_1,M_2) \leq 0\}}\right\}.
\end{equation*}
If $p_{M_1>M_2}<\alpha$, then model $M_1$ is significantly better than $M_2$ in terms of accuracy, at the significance level $\alpha$.

\section{Results}
\label{case_studies}

\subsection{Model comparison and selection}

To select the most suitable modeling approach for the case studies, we compared the performance of the Lasso Logistic regression (LL), Ridge Logistic regression (RL), and Random Forest (RF) models described in Sections~\ref{LogReg} and \ref{RF}. For reference, we also considered the dummy classifier obtained by classifying according to the Majority Class (MC). The models were trained on $70\pct$ of the samples and evaluated on held-out test sets comprising the remaining $30\pct$ for each case study. Performance was assessed using accuracy, with uncertainty quantified via percentile bootstrap confidence intervals and bootstrap one-sided hypothesis tests with $B=100,000$ replicates (see Section~\ref{sec:modcomp}). Tables~\ref{tab:cs1}--\ref{tab:cs3} summarize the evaluations for the competing models.

\begin{table}[!ht]
\centering
\caption{\small Model performance metrics in Case study 1. For each model trained on the training subset, the accuracy (Acc), area under the curve (AUC), and macro $\mathcal{F}_1$ metrics are computed on the testing subset. $\mathrm{CI}(\mathrm{RF},M)$ stands for the percentile bootstrap 95\pct{} confidence interval for $\mathrm{Acc}(\mathrm{RF})-\mathrm{Acc}(M)$ and $p_{\mathrm{RF}>M}$ is the (one-sided) $p$-value for $H_0\colon\mathrm{Acc}(\mathrm{RF})=\mathrm{Acc}(M)$ vs. $H_1\colon\mathrm{Acc}(\mathrm{RF})>\mathrm{Acc}(M)$. Similarly, $\mathrm{CI}(M,\mathrm{MC})$ stands for the percentile bootstrap 95\pct{} confidence interval for $\mathrm{Acc}(M)-\mathrm{Acc}(\mathrm{MC})$ and $p_{M>\mathrm{MC}}$ is the $p$-value for $H_0\colon\mathrm{Acc}(M)=\mathrm{Acc}(\mathrm{MC})$ vs. $H_1\colon\mathrm{Acc}(M)>\mathrm{Acc}(\mathrm{MC})$.}
\label{tab:cs1}
\small
\setlength{\tabcolsep}{4pt}
\begin{tabular}{lccccccc}
\toprule
\textbf{Model ($M$)} & \textbf{Acc} & \textbf{AUC} & $\boldsymbol{\mathcal{F}_1}$ & $\mathbf{CI(RF,\it{M})}$ & $\boldsymbol{p_{\mathrm{RF}>M}}$ & $\mathbf{CI(\it{M}\mathbf{,MC})}$ & $\boldsymbol{p_{M>\mathrm{MC}}}$ \\ \midrule
Ridge Logistic (RL) & 0.6087 & 0.6513 & 0.6084 & $(-0.0059, 0.0692)$ & 0.0506 & $(0.0257, 0.1482)$ & 0.0027 \\
Lasso Logistic (LL) & 0.6028 & 0.6692 & 0.6023 & $(-0.0040, 0.0791)$ & 0.0404 & $(0.0198, 0.1403)$ & 0.0050 \\
Random Forest (RF) & 0.6403 & 0.6694 & 0.6396 & --- & --- & $(0.0593, 0.1779)$ & 0.0001 \\
Majority Class (MC) & 0.5217 & 0.5000 & 0.3429 & --- & --- & --- & --- \\
\bottomrule
\end{tabular}
\end{table}

\begin{table}[!ht]
\centering
\caption{\small Model performance metrics in Case study 2. One-sided bootstrap tests and 95\pct{} confidence intervals. The description of Table~\ref{tab:cs1} applies.}
\label{tab:cs2}
\small
\setlength{\tabcolsep}{4pt}
\begin{tabular}{lccccccc}
\toprule
\textbf{Model ($M$)} & \textbf{Acc} & \textbf{AUC} & $\boldsymbol{\mathcal{F}_1}$ & $\mathbf{CI(RF,\it{M})}$ & $\boldsymbol{p_{\mathrm{RF}>M}}$ & $\mathbf{CI(\it{M}\mathbf{,MC})}$ & $\boldsymbol{p_{M>\mathrm{MC}}}$ \\ \midrule
Ridge Logistic (RL) & 0.5979 & 0.6513 & 0.5908 & $(-0.0214, 0.0697)$ & 0.1599 & $(-0.0080, 0.1206)$ & 0.0485 \\
Lasso Logistic (LL) & 0.5898 & 0.6450 & 0.5850 & $(-0.0161, 0.0804)$ & 0.1014 & $(-0.0188, 0.1153)$ & 0.0834 \\
Random Forest (RF) & 0.6220 & 0.6650 & 0.6188 & --- & --- & $(0.0134, 0.1475)$ & 0.0110 \\
Majority Class (MC) & 0.5416 & 0.5000 & 0.3513 & --- & --- & --- & --- \\
\bottomrule
\end{tabular}
\end{table}

\begin{table}[!ht]
\centering
\caption{\small Model performance metrics in Case study 3. The description of Table~\ref{tab:cs1} applies.}
\label{tab:cs3}
\small
\setlength{\tabcolsep}{4pt}
\begin{tabular}{lccccccc}
\toprule
\textbf{Model ($M$)} & \textbf{Acc} & \textbf{AUC} & $\boldsymbol{\mathcal{F}_1}$ & $\mathbf{CI(RF,\it{M})}$ & $\boldsymbol{p_{\mathrm{RF}>M}}$ & $\mathbf{CI(\it{M}\mathbf{,MC})}$ & $\boldsymbol{p_{M>\mathrm{MC}}}$ \\ \midrule
Ridge Logistic (RL) & 0.5847 & 0.6020 & 0.4764 & $(-0.0487, 0.0441)$ & 0.5571 & $(0.0023, 0.0626)$ & 0.0172 \\
Lasso Logistic (LL) & 0.5731 & 0.6098 & 0.4888 & $(-0.0325, 0.0534)$ & 0.3546 & $(-0.0162, 0.0580)$ & 0.1410 \\
Random Forest (RF) & 0.5824 & 0.6088 & 0.5560 & --- & --- & $(-0.0209, 0.0835)$ & 0.1375 \\
Majority Class (MC) & 0.5522 & 0.5000 & 0.3558 & --- & --- & --- & --- \\
\bottomrule
\end{tabular}
\end{table}

The results for the three case studies show a gradient in the strength of the signal. In Case study 1, RL, LL, and RF all offer mild but significant improvements over MC at significance level $\alpha=5\pct$. In Case study 2, RF ($p_{\mathrm{RF}>\mathrm{MC}}=0.011$) and, at the boundary, RL ($p_{\mathrm{RL}>\mathrm{MC}}=0.049$) improve significantly over MC, while LL does not; the interval reported for RL vs.\ MC still contains zero, since the confidence intervals in Tables~\ref{tab:cs1}--\ref{tab:cs3} are two-sided whereas the $p$-values test a one-sided alternative. In Case study 3, only RL attains a---modest---significant improvement ($p_{\mathrm{RL}>\mathrm{MC}}=0.017$). This connects with the anticipated challenging classification problem revealed by Figure~\ref{fig:male_female_ranges}. No method dominates in terms of point estimates: RF attains the best held-out accuracy in Case studies 1 and 2, and RL in Case study 3. The accuracy of RL is, moreover, competitive with that of RF throughout: the bootstrap confidence intervals for the accuracy difference (Tables~\ref{tab:cs1}--\ref{tab:cs3}) show that RF's advantage is at most modest---between $-0.6$ and $6.9$ accuracy points in Case study 1 ($p_{\mathrm{RF}>\mathrm{RL}}=0.051$)---and compatible with zero in Case studies 2 and 3 ($p_{\mathrm{RF}>\mathrm{RL}}=0.160$ and $0.557$, respectively).

In light of these considerations, ridge logistic regression was adopted as the final model for the three case studies: its predictive performance is competitive with that of the best-performing alternative, and it is the only model among those evaluated that offers interpretability and uncertainty quantification on its coefficients.

Following the model selection discussion, the Ridge Logistic (RL) regression model was applied to the full dataset considered in each case study; the selected penalties were $\hat{\lambda}_{1\mathrm{SE}} = 0.106$, $0.157$, and $0.288$ for Case studies 1--3, respectively (in the parametrization of Eq.~(\ref{eq:eln})). Table~\ref{tab:finalridge} gives the summary of the apparent performance metrics (i.e., computed on the same arias used to fit the model, and hence slightly optimistic) for this model. The best accuracies are obtained for Case studies 1 and 2.\nowidow[3]

\begin{table}[!ht]
\centering
\caption{\small Apparent performance metrics for Ridge Logistic (RL) on the datasets for the three case studies.}
\label{tab:finalridge}
\small
\begin{tabular}{cccc}
\toprule
\textbf{Case study} & \textbf{Acc} & \textbf{AUC} & $\boldsymbol{\mathcal{F}_1}$ \\ \midrule
1 & 0.6278 & 0.6961 & 0.6266 \\
2 & 0.6303 & 0.6955 & 0.6223 \\
3 & 0.6079 & 0.6495 & 0.5625 \\
\bottomrule
\end{tabular}
\end{table}

\subsection{Case study 1: classification of arias by the characters' gender as defined by the opera libretto}
\label{Experiment_1}

Case study 1 examines whether composers adjusted their writing for soprano roles according to the gender assigned to a character in the libretto (i.e., feminine vs. masculine), regardless of the performer's biological sex. The accuracy attained on the held-out arias, although modest, is significantly above the majority-class baseline ($p=0.003$; Table~\ref{tab:cs1}): the musical features carry genuine information on the characters' gender, indicating that this gender was relevant for composers when crafting the melodic lines. The relevance of feature coefficients for this model can be observed in Figure~\ref{fig:Exp1_coefs}.

\begin{figure}[!h]
\centering
\includegraphics[width=0.75\linewidth]{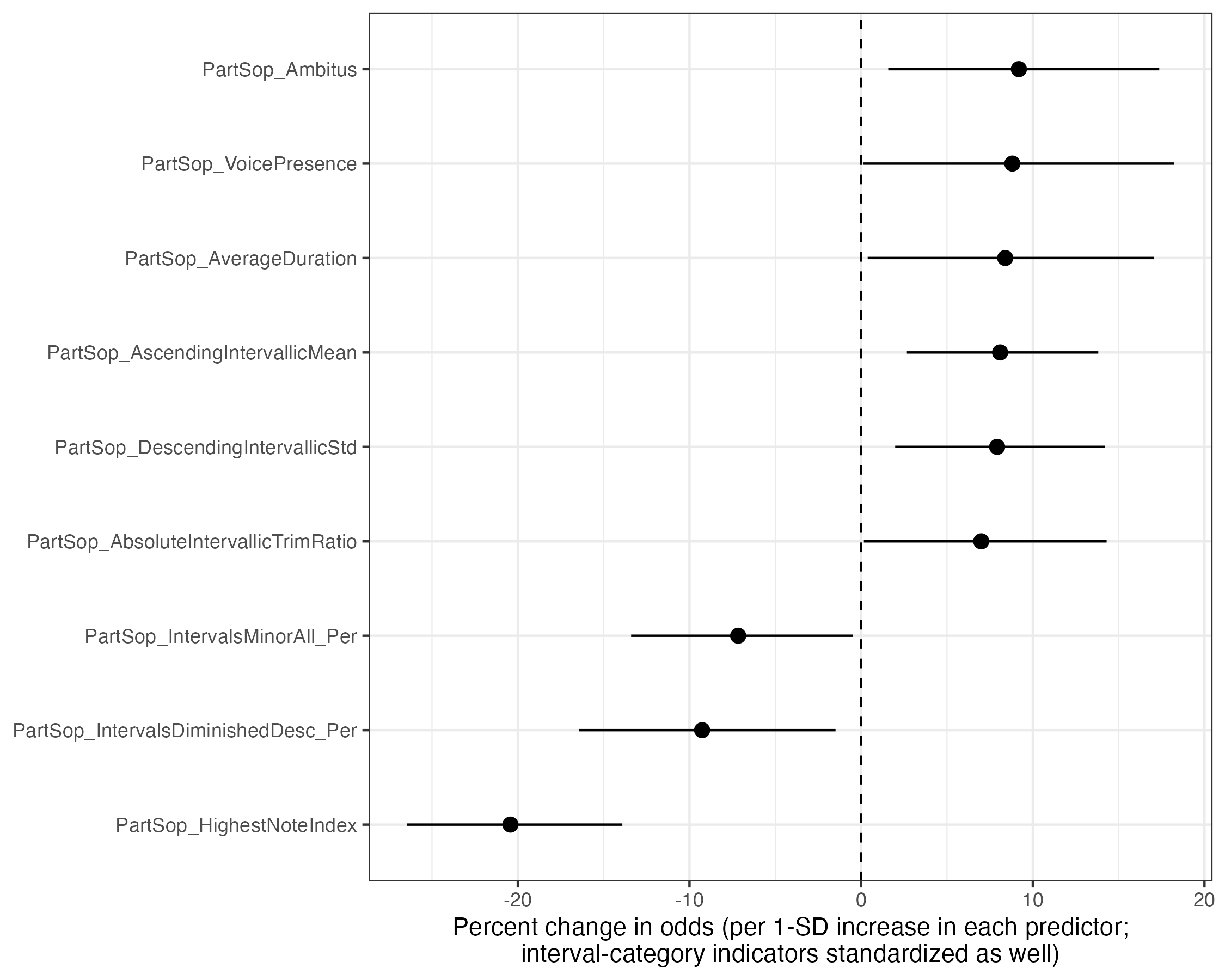}
\caption{\small Case study 1: ridge logistic regression coefficients.
Estimated coefficients and their Bonferroni-corrected asymptotic confidence intervals from Eq.~(\ref{eq:cis}). The estimated coefficients from Eq.~(\ref{eq:eln}) are shown as $(\exp\{\hat{\beta}_{\hat{\lambda}_{\mathrm{1SE}},j}\}-1)\times 100$ so they are interpretable as a percentage change in the odds of $Y=1$ vs. $Y=0$. Since all predictors have been standardized, the percent change in odds is interpreted per standard-deviation increase of each predictor. Only estimated coefficients that are significantly different from zero (i.e., whose corrected confidence interval excludes zero) are shown.}
\label{fig:Exp1_coefs}
\end{figure}

\pagebreak

For the classification of feminine characters, we observe that the most relevant coefficient in our model is \texttt{PartSop\_HighestNoteIndex}. For the sake of economy, in all subsequent feature names, we remove the opening \texttt{PartSop\_}. Definitions of the features mentioned throughout the study are given in Section~\ref{app:feature_definitions}; they all refer to the soprano part only. The prominence of this coefficient indicates that the pitch of the highest note was characteristic of feminine-character arias. In particular, for each standard-deviation increase in the (standardized) \texttt{HighestNoteIndex}, the odds of being a masculine character over feminine decrease by 20.43\pct{} (CI: 13.91\pct, 26.46\pct). Beyond the extreme cases of feminine characters reaching a G6 (e.g., Cleofide's ``Se mai turbo il tuo riposo'', from M.~Mortellari's \textit{Alessandro nell'Indie}, 1778), arias surpassing a D6 are invariably written for feminine characters. By the same token, some arias for masculine soprano characters only reach a B4 as the highest note (e.g., Artabano's ``Amalo, e se al tuo sguardo'', from J.~A.~Hasse's \textit{Artaserse}, 1730), potentially extending down to an E$\flat$3 in the low part of the register (as in Timante's ``Prudente mi chiedi?'' from J.~A.~Hasse's 1748 \textit{Demofoonte}). Yet, these are atypical, outlier cases that do not serve to unequivocally characterize the soprano vocal lines as belonging to either feminine or masculine characters. It should also be noted that \texttt{HighestNoteIndex} captures only the maximum pitch reached in an aria and should not be interpreted as a direct measure of tessitura: two arias may share the same highest note while differing substantially in their overall distribution of pitches. Future work incorporating explicit tessitura descriptors would therefore provide a more refined characterization of vocal writing.

The model also indicates that masculine characters tend to be portrayed, musically speaking, through generally wider vocal ambitus. For each standard-deviation increase in \texttt{Ambitus}, the odds of being a masculine character over feminine increase by 9.19\pct{} (CI: 1.58\pct, 17.37\pct). This tendency is illustrated by, for instance, Enea's ``Se resto sul lido'', from N.~Jommelli's 1747 \textit{Didone abbandonata}, which features a vocal line spanning 29 semitones---i.e., two octaves plus a perfect fourth---from A3 to D6. At the other end of the spectrum, we find, for instance, N.~Piccinni's ``Se troppo crede al ciglio'', for the character of Cleofide in his \textit{Alessandro nell'Indie} (1774), with a vocal line spanning just an octave, from A4 to A5.

The relevance of feminine characters' highest notes is more than 11 points higher than that of the second most relevant feature on the feminine side, \texttt{IntervalsDiminishedDesc\_Per}, indicating the proportion of diminished descending intervals in the arias. In this regard, the odds of being a masculine character over feminine in this model decrease by 9.26\pct{} (CI: 1.49\pct, 16.42\pct) per standard-deviation increase of this feature. This tendency is compatible with traditional associations between diminished intervals and heightened expressive tension, although the present analysis cannot establish the compositional motivation behind this pattern. Feminine characters also appear to sing more minor intervals (\texttt{IntervalsMinorAll\_Per}; 7.16\pct{} decrease in the masculine odds per standard-deviation increase), particularly minor seconds---i.e., diatonic semitones.

This can be observed through the contraposition between two arias from the same opera version: Dircea's ``Se tutti i mali miei'' and Timante's ``Se ardire e speranza'', both from A.~Caldara's \textit{Demofoonte} (1733). In the \textit{prima donna}'s aria, the profusion of diminished intervals---including fifths and thirds, C5--F$\sharp$4 and D$\flat$5--B4 in Figure~\ref{fig:Caldara_SeTutti}; in green---is clear, as well as the presence of semitones (F5--E5, F$\sharp$4--G4, C5--D$\flat$5, and B4--C5; in orange), while there is no augmented interval throughout. In the masculine protagonist's piece, on the contrary, these feminine markers are absent: the line features an augmented second in m. 35 (Figure~\ref{fig:Caldara_SeArdire}; in green), further followed by two major seconds (A4--G4 and G4--F4) and two major thirds (F4--A4 and D5--B$\flat$4; in orange).

The two excerpts from Caldara's opera are also illustrative of the fact that the vocal lines written for feminine characters tend to have fewer leaps than their masculine counterparts, a tendency the model captures through the interval-size features discussed next. As shown in Figure~\ref{fig:Caldara_SeTutti}, the only leaps occurring within an uninterrupted line---i.e., without an intervening rest---are the diminished fifth C5--F$\sharp$4 and the diminished third D$\flat$5--B4 discussed above. They occur, furthermore, across different syllables. In the case of Timante's line (Figure~\ref{fig:Caldara_SeArdire}), on the contrary, there are also leaps within a single, uninterrupted melisma: a minor sixth A4--F5, a perfect fourth A4--D5, and the aforementioned major thirds.\nowidow

\begin{figure}[!h]
\centering
\includegraphics[width=0.9\linewidth]{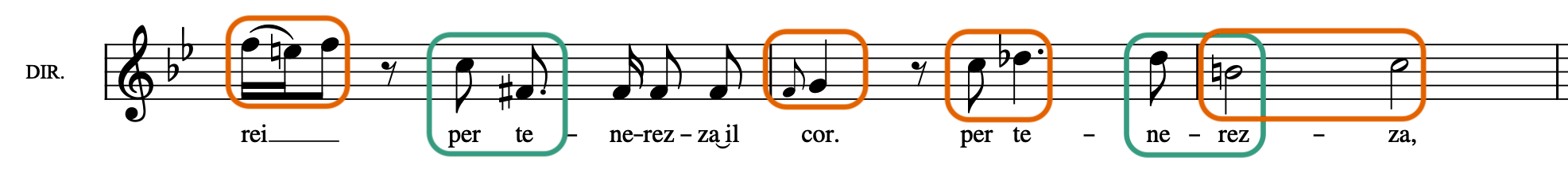}
\caption{\small Diminished and minor intervals in an aria for a feminine character.
Antonio Caldara, \textit{Demofoonte} (1733), ``Se tutti i mali miei,'' mm. 14--16, vocal line. Legend: green = diminished intervals; orange = minor intervals (seconds).}
\label{fig:Caldara_SeTutti}
\end{figure}

\begin{figure}[!h]
\centering
\includegraphics[width=0.9\linewidth]{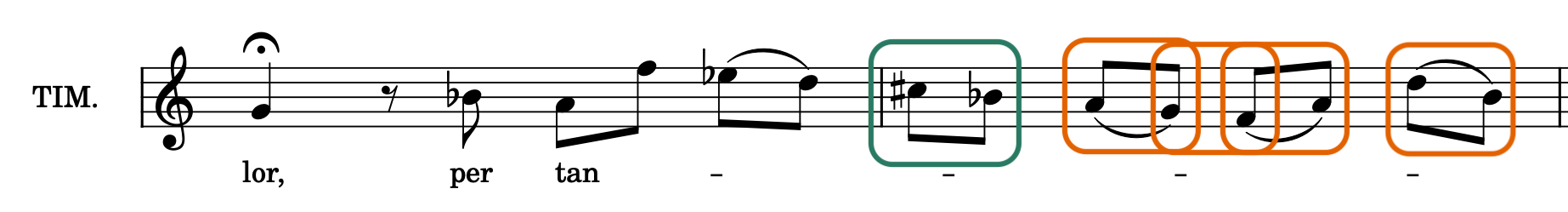}
\caption{\small Absence of the feminine interval markers in an aria for a masculine character.
Antonio Caldara, \textit{Demofoonte} (1733), ``Se ardire e speranza,'' mm. 34--35, vocal line. Legend: green = augmented intervals; orange = major intervals (seconds and thirds).}
\label{fig:Caldara_SeArdire}
\end{figure}

Complementarily, the model indicates that the odds of being a masculine character increase with the average size of the ascending intervals, in semitones (\texttt{AscendingIntervallicMean}; 8.10\pct; CI: 2.66\pct, 13.81\pct). The most extreme case is J.~Ch.~Bach's ``Oh dio! La man mi trema'' from his 1762 setting of \textit{Alessandro nell'Indie}, in which the average ascending interval is larger than an augmented fourth, with instances as large as an octave plus a perfect fifth (D4--A5 in m. 31). At the opposite end, one finds G.~Insanguine's ``Son regina e sono amante'', written for the title character of Didone in his 1772 setting of the homonymous libretto, or L.~Gatti's ``Se il ciel mi divide,'' crafted for Cleofide, the feminine protagonist of \textit{Alessandro nell'Indie}. In both cases---as in many others---the average ascending interval nears the major second, indicating a preponderance of stepwise-motion writing. In fact, approximately 58\pct{} of the arias in which the average ascending interval is smaller than a minor third are for feminine characters, while the ratio decreases to about 47.5\pct{} when the average ascending interval increases from a minor to a major third---that is, with an increase of just one~semitone.

Intervals in masculine-character arias furthermore show a greater variability in size, particularly in the descending direction (\texttt{DescendingIntervallicStd}; 7.92\pct; CI: 1.98\pct, 14.20\pct), meaning they can potentially be of any size. The relative influence of the most extreme intervals on the intervallic mean is likewise significant (\texttt{AbsoluteIntervallicTrimRatio}; 7.00\pct; CI: 0.16\pct, 14.30\pct). Indeed, there are cases of masculine characters performing leaps up to two octaves, as in D.~Perez's ``Destrier, che all'armi usato'' from his \textit{Alessandro} (1755), written for the character of Poro. This large leap occurs within a single melisma, between mm. 36 and 37, G$\sharp$3--G$\sharp$5. Interestingly, too, the smallest maximum ascending interval in the corpus is a diminished fifth and, as expected, appears in an aria for a feminine character---namely Erissena's ``Chi vive amante sai che delira'' from M.~Mortellari's \textit{Alessandro nell'Indie} (1778), whose largest descending leap is in turn only a minor sixth. This is not necessarily related to the ambitus: Erissena's line encompasses an octave plus a diminished fifth (from F$\sharp$4 to C6), which, as discussed above, is average for a feminine character. Mortellari's spare use of large leaps, then, appears deliberate rather than forced by limitations in registral scope.

According to our model, masculine characters are musically associated with intense vocal lines in terms of endurance: the coefficients of the ridge logistic regression indicate a ca.~8--9\pct{} increase in the odds of being a masculine character if the vocal presence (\texttt{VoicePresence}; 8.81\pct) and/or the average duration of the notes (\texttt{AverageDuration}; 8.39\pct) increase. For instance, Poro's ``Se mai pi\`{u} sar\`{o} geloso'' from A.~Sacchini's 1768 \textit{Alessandro} barely has two measures at the end in which the orchestra plays alone, while the rest of the cavatina is dominated by the vocal line, as reflected in the high \texttt{VoicePresence} (defined as the number of measures in which the soprano part sings divided by the total number of measures of the piece). Complementarily, some arias assigned to feminine characters present very short note values, with an average around a sixteenth note, as is the case of Erissena's ``Chi vive amante sai che delira'' from L.~Ko\v{z}eluch's \textit{Alessandro} (1769).

In sum, our model shows that masculine characters have an increased vocal presence and longer note values throughout the arias, while feminine characters seem to be characterized by consistently higher notes and smaller intervals.

\subsection{Case study 2: classification by characters' gender of arias premiered by sopranos of the matching sex}

Case study 2 repeats the classification by characters' gender, restricted to arias in which the character's gender aligns with the singer's biological sex. The model performs about as well as in Case study 1 (see Table~\ref{tab:finalridge}) and, on the held-out arias, RL remains significant against the majority-class baseline, if only at the boundary ($p=0.049$; Table~\ref{tab:cs2}), with RF clearly significant ($p=0.011$), indicating that the gendered signal detected there is not an artifact of cross-cast arias: it persists when attention is restricted to arias in which character and singer align. Whether composers additionally adjusted their writing to the singer's sex must therefore be read from the shifts in the individual features, to which we now turn.

As 76.40\pct{} of the arias for masculine characters were written for male singers (see Table~\ref{tab:distribution_arias}) and 70.68\pct{} of the feminine pieces were written for female singers, it is not surprising that most of the features that were significant for the classification in Case study 1 are also relevant in Case study 2 (Figure~\ref{fig:Exp2_coefs}). However, even within those features some telling differences emerge:
\begin{itemize}
    \item None of the features is as strong a classifier as the highest notes were for feminine characters in Case study 1. This means that, even though the odds of being either a masculine or a feminine character depending on any feature's increase are lower, it is the combination of the various parameters that helps in the classification.
    \item \texttt{HighestNoteIndex} remains the most important feature for the classification of feminine characters, with the odds of being masculine decreasing by 15.42\pct{} (CI: 8.99\pct, 21.40\pct) for each standard-deviation increase in the highest pitch of the aria.
    \item The model indicates that \texttt{Ambitus} (6.95\pct{} odds increase vs. 9.19\pct{} in Case study 1; see Table~\ref{tab:male_features}) may be less relevant for classifying masculine arias when the singer's biological sex is taken into account, while \texttt{VoicePresence} and \texttt{AverageDuration} lose their significance. The latter two do not merely fall short of the significance threshold: their estimated effects also shrink substantially, from 8.81\pct{} to 5.48\pct{} and from 8.39\pct{} to 2.83\pct, respectively.
    \item In contrast, \texttt{Descending\-IntervallicStd} (7.94\pct{} vs. 7.92\pct) and \texttt{AbsoluteIntervallicTrimRatio} (7.92\pct{} vs. 7.00\pct) retain their importance, \texttt{LargestSemitonesAsc} (7.52\pct) and \texttt{TrimmedAbsoluteIntervallicStd} (4.88\pct) emerge as significant, and the relevance of \texttt{AscendingIntervallicMean} decreases (6.33\pct{} vs. 8.10\pct). In fact, \texttt{Descending\-IntervallicStd} and \texttt{AbsoluteIntervallicTrimRatio} are now the clearest classifying features for masculine arias originally written for castrati.
    \item Once again, the model shows that a marked presence of minor intervals helps characterize feminine characters' vocal lines, while the diminished intervals lose their significance here. The feminine class is now also signaled by \texttt{LargestSemitonesDesc} (6.33\pct), i.e., by smaller largest descending leaps (this feature measures descending intervals as negative values, so an increase means a reduction in interval size).
\end{itemize}
\begin{table}[!ht]
\centering
\caption{\small Odds percent increase for masculine/male parts. All the features significantly associated with the masculine/male class in at least one of the models for Case studies 1, 2, and 3. The numbers reported correspond to $(\exp\{\hat{\beta}_{\hat{\lambda}_{\mathrm{1SE}},j}\}-1)\times 100$ in Figures~\ref{fig:Exp1_coefs}, ~\ref{fig:Exp2_coefs}, and \ref{fig:Exp3_coefs}. Empty cells indicate features whose coefficient is not significantly different from zero in that case study.}
\label{tab:male_features}
\small
\begin{tabular}{lccc}
\toprule
\textbf{Feature} & \textbf{Case study 1} & \textbf{Case study 2} & \textbf{Case study 3} \\ \midrule
\texttt{Ambitus} & 9.19 & 6.95 &  \\
\texttt{VoicePresence} & 8.81 & &  \\
\texttt{AverageDuration} & 8.39 & &  \\
\texttt{AscendingIntervallicMean} & 8.10 & 6.33 & 4.29 \\
\texttt{DescendingIntervallicStd} & 7.92 & 7.94 & 5.49 \\
\texttt{AbsoluteIntervallicTrimRatio} & 7.00 & 7.92 & 4.67 \\
\texttt{LargestSemitonesAsc} & & 7.52 & 5.80 \\
\texttt{TrimmedAbsoluteIntervallicStd} & & 4.88 &  \\
\texttt{IntervalsBeyondOctaveAsc\_Per} & & & 4.28 \\
\bottomrule
\end{tabular}
\end{table}
\begin{figure}[!h]
\centering
\includegraphics[width=0.75\linewidth]{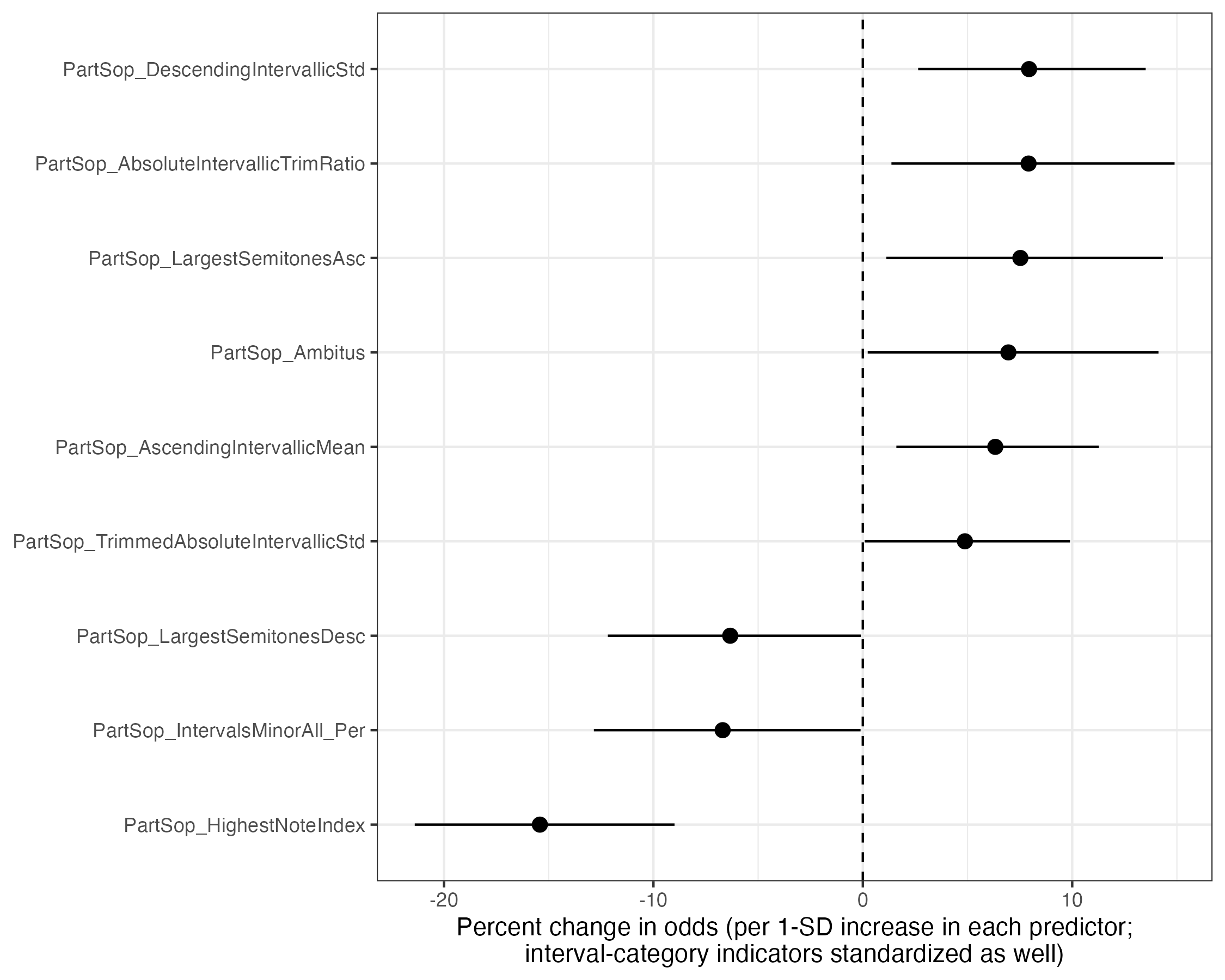}
\caption{\small Case study 2: ridge logistic regression coefficients.
The description of Figure~\ref{fig:Exp1_coefs} applies.}
\label{fig:Exp2_coefs}
\end{figure}

In relation to the fourth point in the above enumeration, masculine characters---as sung by male singers---are, according to our model, now more clearly associated with large ascending leaps, in contrast with their feminine counterparts: the odds of being a masculine character increase by 7.52\pct{} (CI: 1.12\pct, 14.33\pct) per standard-deviation increase in the largest ascending interval of the aria (\texttt{LargestSemitonesAsc}). An example of this is the leaping melodic line in ``Destrier, che all'armi usato'' from G.~M.~Abos' \textit{Alessandro nell'Indie} (1747). Measures 78--86, for instance, are characterized by a chain of ninths (E5--D4), tenths (A3--C5), thirteenths (C4--A5), and even fourteenths (A5--B3) (Figure~\ref{fig:Abos_Destrier}).\nowidow[3]

Importantly, the average duration of the aria's notes (\texttt{AverageDuration}) is no longer relevant in the classification of masculine vocal lines, its estimated effect dropping to a third of its Case study 1 value. This attenuation suggests that the association between masculine characters and longer note values observed in Case study 1 was plausibly influenced by the occasional presence of a female soprano voice premiering them: when the masculine arias considered are sung exclusively by castrati, the feature loses its preponderance.

\begin{figure}[!h]
\centering
\includegraphics[width=0.9\linewidth]{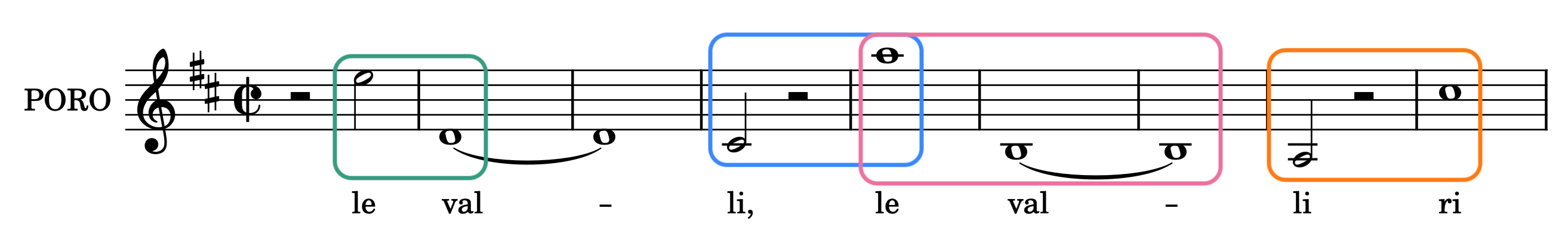}
\caption{\small Chain of intervals larger than the octave in an aria for a masculine character.
Girolamo Matteo Abos, \textit{Alessandro nell'Indie} (1747), ``Destrier, che all'armi usato,'' mm. 78--86, vocal line. Legend: green = major 9th; orange = minor 10th; blue = major 13th; pink = minor 14th.}
\label{fig:Abos_Destrier}
\end{figure}

The vocal presence (\texttt{VoicePresence}) likewise loses its significance in this case study: even though the arias for female performers might be long, there could be extended passages in which the vocal part is silent through rests. A case in point is ``Ombra dell'idol mio'' from D.~Perez's \textit{Alessandro} (1752). Sung by the feminine protagonist Cleofide and written for soprano Caterina Visconti, the vocal line sounds in less than half of the aria (specifically, 46.15\pct). Similarly extreme cases include ``Va crescendo il mio tormento'' from N.~Porpora's \textit{Didone abbandonata} (1725), crafted for the character of Didone in Marianna Benti's voice. Besides the cavatina ``Se mai pi\`{u} sar\`{o} geloso'' by A.~Sacchini (from his \textit{Alessandro}, 1768), discussed above, at the opposite end we find ``Prudente mi chiedi?'' from N.~Jommelli's 1764 \textit{Demofoonte}, in which the character of the masculine protagonist, Timante (through singer Giuseppe Aprile), is present in almost 98\pct{} of the measures.

\subsection{Case study 3: classification of arias by singers' biological sex}

Case study 3 classifies arias by the premiering singer's biological sex (male sopranos being predominantly castrati in this corpus). It should be borne in mind that the two targets are far from independent: in 1,239 of these 1,436 arias (86.28\pct) the singer's sex coincides with the character's gender, so this case study cannot fully disentangle the one from the other. As shown in Table~\ref{tab:finalridge}, the decrease in the model's performance indicates that composition was determined more by the characters' gender than by the biological sex of the singers. On the held-out arias, RL yields a modest but significant improvement over the majority-class baseline ($p=0.017$; Table~\ref{tab:cs3}), while LL and RF do not: the singers' sex leaves a detectable, yet markedly weaker, trace in the notation than the characters' gender. The coefficients examined below are therefore to be read as significant associations between individual features and the singers' sex (Figure~\ref{fig:Exp3_coefs} displays only coefficients significantly different from zero), rather than as evidence of a strong overall predictive signal.

As expected, some of the features that emerge as most relevant for classification in our model are similar to those detected in Case study 2 (and, by extension, in Case study 1 as well) (Figure~\ref{fig:Exp3_coefs}). More specifically, in the classification of female soprano voices, again the highest pitch (\texttt{HighestNoteIndex}) appears to be most relevant, although less so in comparison with the results obtained in the previous classification tasks: here, the odds of being a male soprano decrease by 7.78\pct{} per standard-deviation increase (vs. 20.43\pct{} in Case study 1 and 15.42\pct{} in Case study 2; CI in Case study 3: 3.41\pct, 11.95\pct; see Table~\ref{tab:female_features}). In turn, the comparatively marked use of minor intervals that helped characterize feminine parts in Case studies 1 and 2 (\texttt{IntervalsMinorAll\_Per}) is no longer significant when the singers' sex drives the classification. Similarly, our model does not reveal the use of diminished intervals as significant for the classification of female---or male soprano---voices, irrespective of the characters' gender.

In the case of male soprano voices, the model identifies composers' overall intervallic procedures as relevant in distinguishing them from their female counterparts. More specifically, increases in the intervallic averages (\texttt{AscendingIntervallicMean}), in the relative influence of the most extreme intervals on them (\texttt{AbsoluteIntervallicTrimRatio}), and in the intervallic variability (\texttt{DescendingIntervallicStd}) seem to imply an increased probability that the given vocal parts were written with male soprano voices in mind. As already noted, the effects of most of these features are weaker than in Case studies 1 and 2 (see Table~\ref{tab:male_features}). The exception is the proportion of ascending intervals larger than a perfect octave (\texttt{IntervalsBeyondOctaveAsc\_Per}), which, for the first time, the model reveals to be relevant in the classification task. Even though large leaps were discussed already in Case study 2 with reference to, for instance, Figure~\ref{fig:Abos_Destrier}, the model now shows this leap-related feature to be more clearly related to the vocal typology (i.e., female soprano vs. castrato) than to the character's gender.

\begin{figure}[!h]
\centering
\includegraphics[width=0.75\linewidth]{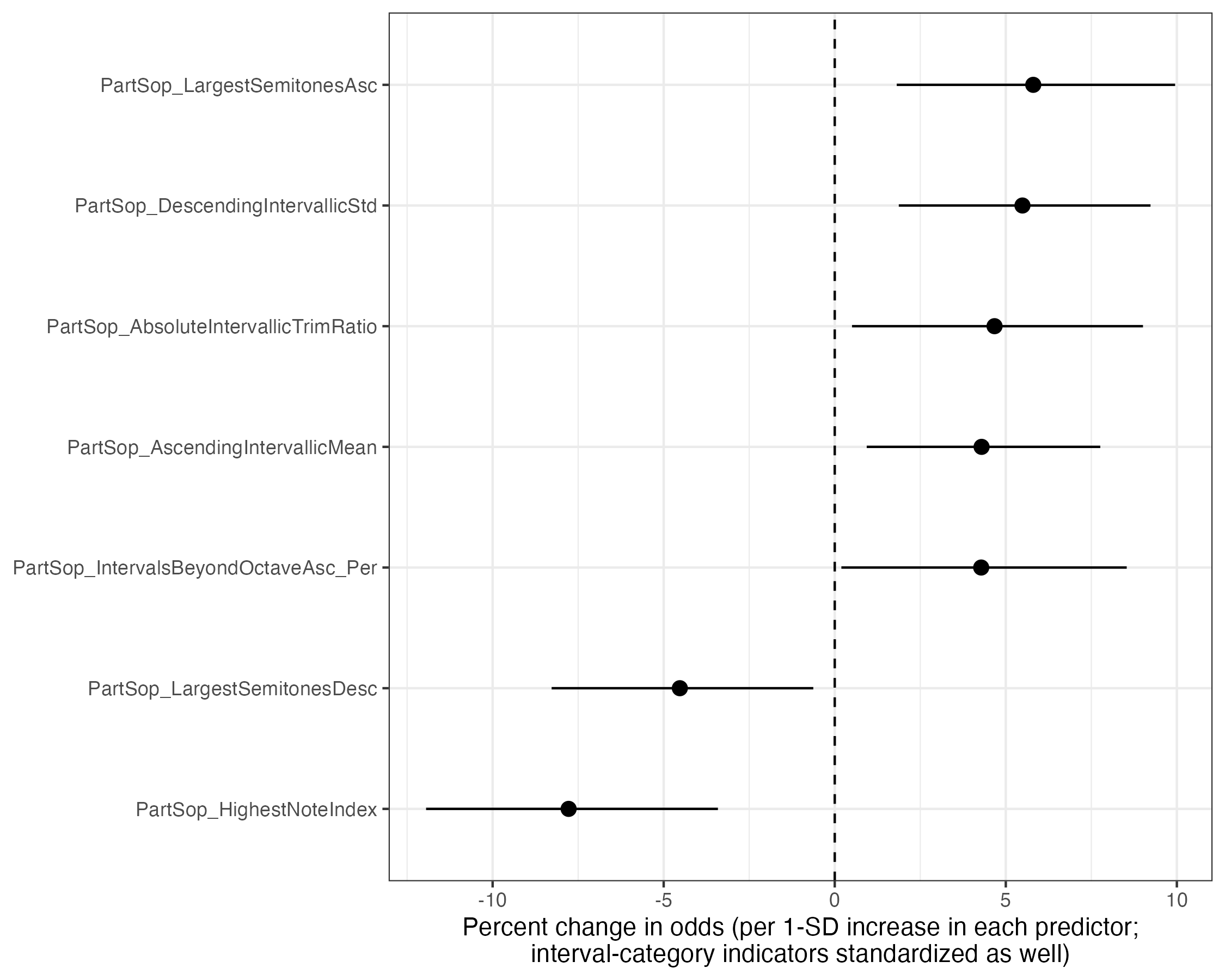}
\caption{\small Case study 3: ridge logistic regression coefficients.
The description of Figure~\ref{fig:Exp1_coefs} applies.}
\label{fig:Exp3_coefs}
\end{figure}

\begin{table}[!ht]
\centering
\caption{\small Odds percent decrease for masculine/male parts. All the features significantly associated with the feminine/female class in at least one of the models for Case studies 1, 2, and 3. The numbers reported correspond to $-(\exp\{\hat{\beta}_{\hat{\lambda}_{\mathrm{1SE}},j}\}-1)\times 100$ in Figures~\ref{fig:Exp1_coefs}, ~\ref{fig:Exp2_coefs}, and \ref{fig:Exp3_coefs}. Empty cells indicate features whose coefficient is not significantly different from zero in that case study.}
\label{tab:female_features}
\small
\begin{tabular}{lccc}
\toprule
\textbf{Feature} & \textbf{Case study 1} & \textbf{Case study 2} & \textbf{Case study 3} \\ \midrule
\texttt{HighestNoteIndex} & 20.43 & 15.42 & 7.78 \\
\texttt{IntervalsDiminishedDesc\_Per} & 9.26 & &  \\
\texttt{IntervalsMinorAll\_Per} & 7.16 & 6.69 &  \\
\texttt{LargestSemitonesDesc} & & 6.33 & 4.53 \\
\bottomrule
\end{tabular}
\end{table}

As in Case study 2, \texttt{LargestSemitonesDesc}---which measures descending intervals as negative values (a descending perfect octave is $-12$, not 12), so that an increase necessarily means a reduction in the interval size ($|-12|>|-7|$)---is relevant for the classification of female voices. Overall, then, the model indicates that male singers were made to sing larger leaps in both directions, the most extreme case among the arias premiered by male sopranos being ``Sperai vicino il lido'' from A.~Ferradini's \textit{Demofoonte} (1759), which was crafted for Filippo Elisi and includes a descending leap of 26 semitones---i.e., two octaves plus a major second (B5--A3 in the middle of a melisma in mm.~53--54).

The excerpts in Figures~\ref{fig:Porpora_ViviSuperbo} and~\ref{fig:Graun_SeMai}---from N.~Porpora's ``Vivi superbo e regna'' (\textit{Didone abbandonata}, 1725) and C.~H.~Graun's ``Se mai turbo il tuo riposo'' (\textit{Alessandro nell'Indie}, 1744)---readily exemplify these contrasts. The perfect and minor intervals marked in the figures describe the intervallic palettes of the two excerpts, although their overall proportions are not significant classifiers in this case study.

\begin{figure}[!h]
\centering
\includegraphics[width=0.9\linewidth]{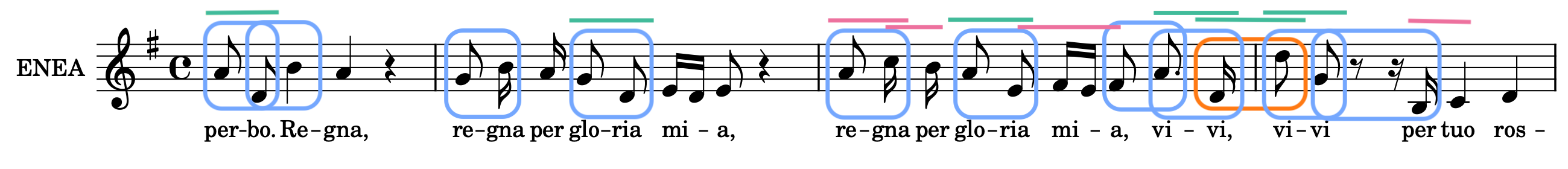}
\caption{\small Vocal writing for a castrato.
Nicola Porpora, \textit{Didone abbandonata} (1725), ``Vivi superbo e regna,'' mm. 11--14, vocal line. Legend: orange = octave intervals or larger; blue = other leaps; green = perfect intervals; pink = minor intervals.}
\label{fig:Porpora_ViviSuperbo}
\end{figure}

\begin{figure}[!h]
\centering
\includegraphics[width=0.9\linewidth]{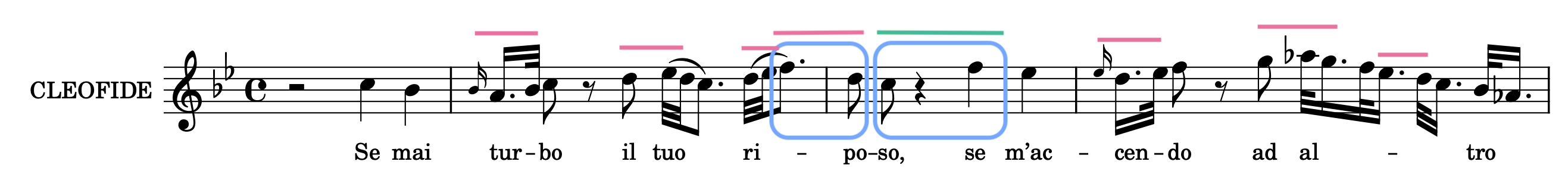}
\caption{\small Vocal writing for a female soprano.
Carl Heinrich Graun, \textit{Alessandro nell'Indie} (1744), ``Se mai turbo il tuo riposo,'' mm. 17--20, vocal line. Legend: blue = leaps; green = perfect intervals; pink = minor intervals.}
\label{fig:Graun_SeMai}
\end{figure}

The feature values for these two arias are given in Table~\ref{tab:2arias}. The only potentially surprising value might be that of the feature \texttt{AbsoluteIntervallicTrimRatio}, which is lower in the case of the aria written for a male soprano. The value differences between the two arias for this feature run against the model's overall direction, but the associated effect is one of the weakest in this case study (a 4.67\pct{} increase in the male odds per standard-deviation increase, the two arias differing by about 1.5 standard deviations in the feature). As expected, and even though the overall ambitus is similar in the two arias, ``Se mai turbo'' reaches the highest pitch (B$\flat$5) and has all its intervals within a major sixth (therefore none beyond the octave).

\begin{table}[!ht]
\centering
\caption{\small Values for classification-relevant numerical features in two contrasting arias.
Nicola Porpora's ``Vivi superbo e regna,'' \textit{Didone abbandonata} (1725), written for castrato
Nicola Grimaldi, and Carl Heinrich Graun's ``Se mai turbo il tuo riposo,'' \textit{Alessandro nell'Indie}
(1744), for soprano Giovanna~Gasparini.}
\label{tab:2arias}
\small
\begin{tabular}{lcc}
\toprule
\textbf{Feature} & \textbf{Porpora (1725), castrato} & \textbf{Graun (1744), female soprano} \\ \midrule
\texttt{LargestSemitonesAsc} & 12 & 9 \\
\texttt{DescendingIntervallicStd} & 2.6889 & 0.9644 \\
\texttt{AbsoluteIntervallicTrimRatio} & 0.0895 & 0.1430 \\
\texttt{IntervalsBeyondOctaveAsc\_Per} & 0 & 0 \\
\texttt{AscendingIntervallicMean} & 4.1786 & 3.0104 \\
\texttt{LargestSemitonesDesc} & $-$12 & $-$8 \\
\texttt{HighestNoteIndex} & 74 & 82 \\
\bottomrule
\end{tabular}
\end{table}

\section{Discussion}

In this work, we have presented the first large-scale, corpus-based study of soprano vocal typology in eighteenth-century Italian opera seria. Several features emerged as the most decisive in the classification tasks performed across all three case studies, all relating to the individual crafting of the vocal line.

The relevant melodic features can be classified into four main categories, as follows:
\begin{itemize}
    \item \textbf{Ambitus or range}. Masculine and feminine characters and singers show a clearly distinctive use of the vocal ambitus, with feminine characters---performed by female singers (Case study 2) or not (Case study 1)---being consistently associated with the highest notes, potentially unsuitable for representing a masculine character. This feature (\texttt{HighestNoteIndex}) is also directly associated with the singer's sex (Case study 3), although there appears to be a more direct correlation with the character's gender. Overall, and mostly due to their more extreme low notes, masculine characters are depicted through a larger ambitus or vocal range (\texttt{Ambitus}), an association that weakens when the singer's sex is targeted, to the point of losing significance in Case study 3. In terms of the fitted models, lowering the bottom of the range by one semitone (at a fixed highest note) increases the masculine odds by 3.08\pct, 2.27\pct, and 1.12\pct{} in the three case studies, whereas raising the top by one semitone (at a fixed ambitus) decreases them by 8.80\pct, 6.39\pct, and 3.19\pct: per semitone, the upper end of the register carries roughly three times more discriminative weight than the lower end. (The Case study 3 figures rest on the \texttt{Ambitus} coefficient that is itself not significant there, and are quoted only for completeness.)

    \item \textbf{Size of the intervals}. Masculine characters---and male singers---show melodic lines with larger intervals, and, therefore, more numerous jumps. This is reflected through various metrics across the three case studies (intervallic averages/ratios, size of the largest leaps, and proportion of ascending leaps larger than the octave). Again, the models performed better when classifying gendered roles than singers' sex.

    When disregarding the most extreme intervals, again, masculine characters tend to be less consistent in their intervallic profile (as shown by higher \texttt{AbsoluteIntervallicTrimRatio} values across the three case studies) than their feminine counterparts. This shows that the increased intervallic average for masculine characters is produced by the presence of potentially---very---large intervals on specific occasions, not by an increased frequency of mid-size leaps. When the character's gender is not targeted in the classification task (Case study 3), again, lower odds in the classification may indicate that male singers received more stable melodies when performing feminine characters.

    \item \textbf{Type of the intervals}. Overall, feminine-character arias in this corpus are associated with a higher proportion of minor intervals (Case studies 1 and 2), and the association between diminished intervals and femininity seems to have been reserved for dramatic depictions (Case study 1) rather than for vocal characteristics (Case study 3). Smaller largest descending leaps (\texttt{LargestSemitonesDesc}), in turn, signal the feminine/female class in Case studies 2 and 3.

    \item \textbf{Vocal presence}. The vocal presence, like the average note duration, is associated with the gender of the character only when the full corpus is considered (Case study 1), and loses its significance when the singer's sex is involved (Case studies 2 and 3). Thus, the relative presence of masculine characters' melodic interventions does not reflect a sexed perspective in terms of the singers, even though it has traditionally been hypothesized to be a consequence of castrati's larger thoracic capacities---a hypothesis for which the literature adduces the long-held notes and extended coloraturas demanded by parts written for castrati \citep[p.~24]{seedorf2015}. Our corpus-scale results suggest, instead, that these traits attach primarily to the character rather than to the singer. One possible explanation is that these features reflect differences in dramatic prominence, since leading characters often receive more extended vocal writing. Assessing this hypothesis would require explicitly modeling dramatic rank, which lies beyond the scope of the present study.
\end{itemize}

Assuming that the majority of masculine-character arias were originally written for male singers (and of feminine-character pieces for female sopranos), it is not surprising that a significant number of features coincide among the three case studies, although their hierarchy of importance varies, showing a greater tendency for characters to be associated with ambitus (including the use of extreme notes). When the singer's biological sex is considered, however, their relative importance decreases, and, especially in the case of male classifications in Case studies 2 and 3, it is surpassed by other, interval-related features.

The experimental results thus indicate a gendered melodic representation of characters---and, very secondarily, of vocal types: feminine characters as steadier, yet experiencing more poignant emotions (as potentially reflected in minor and diminished intervals), and masculine characters as more unstable. Arias written for female singers tend to reach higher maximum pitches, although the decrease in this feature's preponderance across all three case studies indicates that it may have been more clearly associated with the characters than with the singers themselves. That is, although vocal typology plays a role in the classification of singers in opera seria, dramaturgical considerations and casting practices may limit the ability to clearly separate gendered vocal characteristics.

Furthermore, the difference in interval usage for either feminine or masculine characters may also be contingent on the emotions associated with them. For instance, in the five libretti with which we have worked---i.e., \textit{Didone abbandonata}, \textit{Alessandro nell'Indie}, \textit{Artaserse}, \textit{Adriano in Siria}, and \textit{Demofoonte}---arias expressive of ``pity'' are split into 165 for feminine characters vs. 113 for masculine ones, whereas ``anger'' arias are distributed between 194 for masculine vs. 31 for feminine characters. The data were obtained from \citet{didonedatabase}, which is based on the method proposed by \citet{TorrentePassions}. Further studies could shed clearer light on this.

However, the overall accuracy of the model in the three case studies (between 0.58 and 0.61 on the held-out arias) indicates that the results must be evaluated with caution, as the two classes (male/masculine vs. female/feminine) are only moderately well classified. As shown in Case study 3, one cannot expect all pieces in the corpus to be perfect examples of all the variable descriptions shown by the models. The results must, therefore, be considered within the complex and variable scenario of opera production in the eighteenth century, in which personal, economic, and political intricacies played a non-negligible role.

Moreover, the fact that the model performed worst in Case study 3 might be due to the vocal constraints of the specific singers premiering the operas, but may also reflect the prioritization of the narrative function showcased in each aria---although these spheres often overlapped. In other words, it might mean that composers were more preoccupied with the feminine vs. masculine traits of the characters (or, more broadly, with the narrative relevance of the character portrayed) than with the vocal constraints of the singers.

A complementary explanation is suggested by the nature of the sources themselves. The distinctiveness of the castrato voice, as described by contemporaries, lay in the quality of the voice itself---power, timbre, breath control---precisely the parameters that musical notation does not record. Improvised ornamentation, by contrast, was no castrato monopoly: it was the shared art of every virtuoso singer, women as much as men \citep{Poriss}, and it equally escapes the notated score (see Section~\ref{sec:intro}). What the notated score could encode was the character; what the theatre experienced as unmistakably different was the singer. Nor was that sound a uniform category: individual castrato voices seem to have differed widely from one another, and period descriptions may reflect particular singers rather than a common type \citep{Sundberg_sopranos}. The comparatively weak notational signal for the singer's sex in Case study 3 is thus not merely a limitation of the model: it is evidence about where the difference between castrati and female sopranos actually resided---in sound, not in~notes.

In all three case studies, more features emerged as relevant in the models for the classification of male/masculine vocal lines (for characters and/or singers): 6 vs. 3 in Case studies 1 and 2, and 5 vs. 2 in Case study 3. This seems to indicate that the vocal lines written for masculine characters (mostly sung by male sopranos, too) were more distinctive than those of feminine roles. The contrast between the classes is one of distribution, not of magnitude---the typical effect sizes are comparable (Case study 1 medians: 8.25 male/masculine vs. 9.26 female/feminine), but the male/masculine signal is spread over more features while the female/feminine one is concentrated in the highest note. When \texttt{HighestNoteIndex} is dropped from the analysis, the feminine odds totals fall to 16.4 (Case study 1) / 13.0 (Case study 2) / 4.5 (Case study 3) against male/masculine totals of 49.4 / 41.5 / 24.5.

This might be related to the fact that, quite frequently, female singers had to perform masculine soprano roles on the one hand and, on the other, that some feminine protagonist roles in Metastasio's dramas were strong women, such as Didone, exhibiting traits of decisiveness---particularly in the first act---conventionally coded as masculine within eighteenth-century dramaturgy. In this regard, note the reduction in arias to be classified in Case study 2, compared to Case study 1 (see the alignment percentages quoted there); the ratio of missing data regarding the premiering singer is similar between the two groups (ca. 14--15\pct).

Furthermore, we may surmise that the use of vocal features would also need to be coherent with the rest of the score. This highlights the holistic nature of operatic composition, in which the vocal attributes interact with broader narrative and musical structures rather than existing in isolation. In this historical context, composers likely designed their works to be adaptable across productions, allowing for a range of singers to perform them irrespective of their biological sex. As such, the models obtained in the case studies presented highlight the complexity and nuance of opera seria~composition.

\section{Conclusion}

Taken together, the three case studies reveal an asymmetry that neither a rigid nor a fully flexible account of opera seria conventions would predict. The bodies were, in practice, interchangeable: the singer's biological sex is only weakly recoverable from the notated music (Case study 3). The writing, however, was gendered: the character's dramatic gender is consistently the better-predicted variable across all models (Case studies 1 and 3). Composers thus encoded gender in the melodic fabric of the arias while writing for a market in which either type of soprano might ultimately perform them. This double condition---flexible casting sustained by gender-coded composition---refines, rather than merely confirms, the received picture of the genre: the flexibility long documented in casting practice did not extend to a neutral, ungendered vocal writing; and, conversely, the gender coding of that writing was dramatic rather than physiological in origin, anchored to the character rather than to the body that sang it.

\section*{Supplementary materials}

The supplementary materials define the \texttt{musif} features referred to throughout the study (Section~\ref{app:feature_definitions}) and list the complete set of 86 features considered before preprocessing (Section~\ref{app:features_list}).

The analyzed data are deposited on Zenodo under CC BY 4.0 (\href{https://doi.org/10.5281/zenodo.21757127}{doi.org/10.5281/zenodo.21757127}) and the analysis code at \href{https://github.com/DIDONEproject/SopranoVoices}{github.com/DIDONEproject/SopranoVoices}.

\section*{Acknowledgments}

This work was supported by the European Research Council (ERC) under the European Union's Horizon 2020 research and innovation programme [grant agreement No.~788986, project DIDONE].


\fi

\ifsupplement

\newpage
\title{Supplementary materials for ``Soprano voices in opera seria: a corpus-based inquiry into eighteenth-century vocal types''}
\setlength{\droptitle}{-1cm}
\predate{}%
\postdate{}%
\date{}

\author{Ana Llorens$^{1,4}$, Eduardo Garc\'ia-Portugu\'es$^{2}$, Carlos Vaquero$^{3}$, and \'Alvaro Torrente$^{1,3}$}
\footnotetext[1]{Department of Musicology, Universidad Complutense de Madrid (Spain).}
\footnotetext[2]{Department of Statistics, Universidad Carlos III de Madrid (Spain).}
\footnotetext[3]{Instituto Complutense de Ciencias Musicales, Madrid (Spain).}
\footnotetext[4]{Corresponding author. e-mail: \href{mailto:allorens@ucm.es}{allorens@ucm.es}.}
\maketitle

\begin{abstract}
	These supplementary materials contain two sections. Section~\ref{app:feature_definitions} gives operational definitions of the musical features referenced in the discussion of results, covering measures of vocal range, intervallic size and variability, interval quality (e.g., major, minor, diminished, augmented), leap frequency, and vocal presence. Section~\ref{app:features_list} lists the complete set of candidate features extracted prior to the preprocessing described in Section~\ref{sec:preproc} of the main text.
\end{abstract}
\begin{flushleft}
	\small\textbf{Keywords:} Castrato; Corpus study; Gender; Logistic regression; Opera seria; Soprano; Statistical learning; Vocal range.
\end{flushleft}

\appendix

\section{Definitions of selected features}
\label{app:feature_definitions}

This appendix defines the \texttt{musif} features mentioned throughout the study. They all refer to the soprano part only, and their names are given with the opening \texttt{PartSop\_} removed.
\begin{itemize}
\setlength{\itemsep}{0pt}\setlength{\parskip}{0pt}
    \item \texttt{AbsoluteIntervallicTrimRatio}: Difference between the \texttt{AbsoluteIntervallicMean} (i.e., average of the absolute interval sizes, in no. of semitones, in the soprano part) and the \texttt{TrimmedAbsoluteIntervallicMean} (i.e., trimmed mean of the absolute interval sizes, discarding 10\pct{} of the most extreme data), divided by the \texttt{AbsoluteIntervallicMean}. Larger values thus indicate a stronger influence of the most extreme intervals on the intervallic mean.
    \item \texttt{Ambitus}: Ambitus of the soprano part, in number of semitones.
    \item \texttt{AscendingIntervallicMean}: Mean of the sizes of the ascending intervals (in no. of semitones).
    \item \texttt{AverageDuration}: Average duration of the note values in the soprano part (quarter note = 1, eighth note = 0.5, etc.).
    \item \texttt{DescendingIntervallicStd}: Standard deviation of the sizes of the descending intervals (in no. of semitones).
    \item \texttt{HighestNoteIndex}: MIDI pitch of the highest note.
    \item \texttt{IntervalsBeyondOctaveAsc\_Per}: Proportion of ascending intervals larger than a perfect octave, relative to all melodic intervals in the~aria.
    \item \texttt{IntervalsDiminishedDesc\_Per}: Proportion of descending diminished intervals, relative to all melodic intervals in the~aria.
    \item \texttt{IntervalsMinorAll\_Per}: Proportion of minor intervals, relative to all melodic intervals in the~aria.
    \item \texttt{LargestSemitonesAsc}: Largest ascending interval across the soprano part, in number of semitones.
    \item \texttt{LargestSemitonesDesc}: Largest descending interval across the soprano part, in number of semitones.
    \item \texttt{TrimmedAbsoluteIntervallicStd}: Standard deviation of the absolute interval sizes (in no. of semitones), discarding 10\pct{} of the most extreme data.
    \item \texttt{VoicePresence}: Ratio between the \texttt{SoundingMeasures} (i.e., number of measures that have at least one note for the soprano part) and the total number of measures of the score.
\end{itemize}

All basic definitions can be found in the \texttt{musif} documentation:
\url{https://musif.didone.eu/Feature_definition.html}.

\section{Features list}
\label{app:features_list}

The features considered in this study, prior to the preprocessing described in Section~\ref{sec:preproc}, are listed~\mbox{below.}

\begin{multicols}{2}
\footnotesize
\begin{itemize}
\setlength{\itemsep}{0pt}\setlength{\parskip}{0pt}
    \item \texttt{AbsoluteIntervallicKurtosis}
    \item \texttt{AbsoluteIntervallicMean}
    \item \texttt{AbsoluteIntervallicSkewness}
    \item \texttt{AbsoluteIntervallicStd}
    \item \texttt{AbsoluteIntervallicTrimDiff}
    \item \texttt{AbsoluteIntervallicTrimRatio}
    \item \texttt{Ambitus}
    \item \texttt{AscendingIntervallicMean}
    \item \texttt{AscendingIntervallicStd}
    \item \texttt{AscendingIntervals\_Per}
    \item \texttt{AverageDuration}
    \item \texttt{Density}
    \item \texttt{DescendingIntervallicMean}
    \item \texttt{DescendingIntervallicStd}
    \item \texttt{DescendingIntervals\_Per}
    \item \texttt{DottedRhythm}
    \item \texttt{DoubleDottedRhythm}
    \item \texttt{DynAbruptness}
    \item \texttt{DynGrad}
    \item \texttt{DynMean}
    \item \texttt{DynMean\_weighted}
    \item \texttt{HighestNoteIndex}
    \item \texttt{IntervallicKurtosis}
    \item \texttt{IntervallicMean}
    \item \texttt{IntervallicSkewness}
    \item \texttt{IntervallicStd}
    \item \texttt{IntervallicTrimDiff}
    \item \texttt{IntervallicTrimRatio}
    \item \texttt{IntervalsAugmentedAll\_Per}
    \item \texttt{IntervalsAugmentedAsc\_Per}
    \item \texttt{IntervalsAugmentedDesc\_Per}
    \item \texttt{IntervalsBeyondOctaveAll\_Per}
    \item \texttt{IntervalsBeyondOctaveAsc\_Per}
    \item \texttt{IntervalsBeyondOctaveDesc\_Per}
    \item \texttt{IntervalsDiminishedAll\_Per}
    \item \texttt{IntervalsDiminishedAsc\_Per}
    \item \texttt{IntervalsDiminishedDesc\_Per}
    \item \texttt{IntervalsDoubleAugmentedAll\_Per}
    \item \texttt{IntervalsDoubleAugmentedAsc\_Per}
    \item \texttt{IntervalsDoubleAugmentedDesc\_Per}
    \item \texttt{IntervalsDoubleDiminishedAll\_Per}
    \item \texttt{IntervalsDoubleDiminishedAsc\_Per}
    \item \texttt{IntervalsDoubleDiminishedDesc\_Per}
    \item \texttt{IntervalsMajorAll\_Per}
    \item \texttt{IntervalsMajorAsc\_Per}
    \item \texttt{IntervalsMajorDesc\_Per}
    \item \texttt{IntervalsMinorAll\_Per}
    \item \texttt{IntervalsMinorAsc\_Per}
    \item \texttt{IntervalsMinorDesc\_Per}
    \item \texttt{IntervalsPerfectAll\_Per}
    \item \texttt{IntervalsPerfectAsc\_Per}
    \item \texttt{IntervalsPerfectDesc\_Per}
    \item \texttt{IntervalsWithinOctaveAll\_Per}
    \item \texttt{IntervalsWithinOctaveAsc\_Per}
    \item \texttt{IntervalsWithinOctaveDesc\_Per}
    \item \texttt{LargestAbsoluteSemitonesAll}
    \item \texttt{LargestAbsoluteSemitonesAsc}
    \item \texttt{LargestAbsoluteSemitonesDesc}
    \item \texttt{LargestIntervalAll}
    \item \texttt{LargestIntervalAsc}
    \item \texttt{LargestSemitonesAll}
    \item \texttt{LargestSemitonesAsc}
    \item \texttt{LargestSemitonesDesc}
    \item \texttt{LeapsAll\_Per}
    \item \texttt{LeapsAsc\_Per}
    \item \texttt{LeapsDesc\_Per}
    \item \texttt{LowestNoteIndex}
    \item \texttt{RhythmInt}
    \item \texttt{SmallestAbsoluteSemitonesAll}
    \item \texttt{SmallestAbsoluteSemitonesAsc}
    \item \texttt{SmallestAbsoluteSemitonesDesc}
    \item \texttt{SmallestIntervalAll}
    \item \texttt{SmallestIntervalDesc}
    \item \texttt{SmallestSemitonesAll}
    \item \texttt{SmallestSemitonesAsc}
    \item \texttt{SmallestSemitonesDesc}
    \item \texttt{SoundingDensity}
    \item \texttt{StepwiseMotionAll\_Per}
    \item \texttt{StepwiseMotionAsc\_Per}
    \item \texttt{StepwiseMotionDesc\_Per}
    \item \texttt{SyllabicRatio}
    \item \texttt{TrimmedAbsoluteIntervallicMean}
    \item \texttt{TrimmedAbsoluteIntervallicStd}
    \item \texttt{TrimmedIntervallicMean}
    \item \texttt{TrimmedIntervallicStd}
    \item \texttt{VoicePresence}
\end{itemize}
\end{multicols}

\fi

\end{document}